\documentclass[3p,preprint,number,sort&compress,times]{elsarticle}
\usepackage{natbib}
\usepackage{amsmath} % provides a handful of options for displaying equations
\usepackage{amsfonts} % extended set of fonts for use in mathematics
\usepackage{bm} 
\newcommand{\olsi}[1]{\,\overline{\!{#1}}} % overline short italic
\usepackage{calc} % for placement of vdots
\usepackage{float}
\usepackage{subcaption} % enables subfigures
\usepackage{hyperref} % to hyperlink all cross-referenced elements
\usepackage[normalem]{ulem}
\usepackage{xcolor}
\usepackage{algorithm}
\usepackage{algpseudocode}

\hypersetup{pdfauthor={Name}} % to suppress annoying (and unimportant) warnings

\usepackage{amsthm}

\begin{document}
    \begin{titlepage}
    \title{{Feedback linearisation of mechanical systems using data-driven models}}
    
    \author[1]{Merijn Floren\corref{cor1}}
    \ead{merijn.floren@kuleuven.be}
    \author[2]{Koen Classens}
    \ead{k.h.j.classens@tue.nl}
    \author[2]{Tom Oomen}
    \ead{t.a.e.oomen@tue.nl}
    \author[1]{Jean-Philippe Noël}
    \ead{jp.noel@kuleuven.be}
    \cortext[cor1]{Corresponding author}
   
    \address[1]{KU Leuven, Department of Mechanical Engineering, Robotics, Automation \& Mechatronics, Celestijnenlaan 300, 3001 Leuven, Belgium}
    \address[2]{Eindhoven University of Technology, Department of Mechanical Engineering, Control Systems Technology Group, PO Box 513, 5600 MB
    Eindhoven, The Netherlands}

    \begin{abstract}
    Linearising the dynamics of nonlinear mechanical systems is an important and open research area. A common approach is feedback linearisation, which is a nonlinear control method that transforms the input-output response of a nonlinear system into an equivalent linear one. The main problem with feedback linearisation is that it requires an accurate first-principles model of the system, which are typically hard to obtain. In this paper, we design an alternative control approach that exploits data-driven models to linearise the input-output response of nonlinear mechanical systems.{Specifically, a model-based reference tracking architecture is developed for nonlinear feedback systems with output nonlinearities. } The overall methodology shows a high degree of performance combined with significant robustness against imperfect modelling and extrapolation. These findings are demonstrated using large set of synthetic experiments conducted on a asymmetric Duffing oscillator and using an experimental prototype of a high-precision motion system.
    \end{abstract}
    
    \begin{keyword}
    feedback linearisation \sep
    nonlinear mechanical systems \sep
    nonlinearity \sep
    data-driven modelling 
    \end{keyword}

\end{titlepage}
    \maketitle
    \section{Introduction} \label{sec:introduction}
Following the rapid development of our modern society, engineering systems are becoming increasingly complex. In mechanics, this complexity translates in most applications in the appearance of nonlinear phenomena. Consider, for instance, the aerospace sector, where the demands for reduced fuel consumption, increased payload and increased flight range are accommodated by designing ever-lighter aircraft structures \cite{dursun}. An inevitable result of this quest for squeezing the structural weight is the activation of pronounced nonlinear effects \cite{noel2017nonlinear}. Nonlinear dynamic behaviours can only be accurately captured by nonlinear models. In terms of control engineering however, it is much more appealing to work with linear models. In order to meet the engineers' desire, research efforts amongst a range of disciplines have focused on linearising nonlinear systems using feedback control.\\

\begin{figure}[!t]
    \centering
    \includegraphics{ 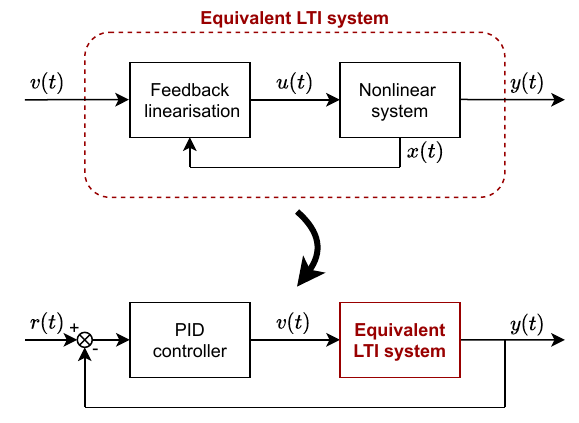}
    \caption{Block diagram illustrating the simplified working principle of feedback linearisation. In a first step, the plant input $u(t)$ is modified such that for the new input $v(t)$ a linear output response is obtained. In a second step, the equivalent LTI system can be integrated in any linear outer-loop control scheme, in this example a classical PID feedback loop with reference tracking.}
    \label{fig:FBLscheme}
\end{figure}

Fig. \ref{fig:FBLscheme} illustrates the general working principle of feedback linearisation for a single-input single-output (SISO) nonlinear system. In essence, the plant input $u(t)$ is modified, through a choice of a nonlinear state feedback control law, such that the input-output (IO) relation between the new outer-loop input $v(t)$ and system output $y(t)$ is linear. This linear relation holds for the entire specified operating range, meaning that the IO response of an equivalent linear time-invariant (LTI) system is obtained. Subsequently, standard linear control techniques can be used to design an outer-loop controller.\\

The commonly used technique for feedback linearisation is the analytical approach based on Lie algebra \cite{sastry,khalil,wagg}. This approach intends to find a direct linear relationship between the system output and a new control input, and eventually transforms the original system into a chain of integrators. In a recent study \cite{huang2020robust}, analytical feedback linearisation was applied in combination with an outer-loop linear quadratic regulator in order to control the desired altitude and attitude levels of a quadrotor. In \cite{gionfra2016combined}, the same approach was applied to linearise the dynamics of a nonlinear wind turbine, after which the obtained LTI system supported the design of an outer-loop model-based controller. A comparison between analytical feedback linearisation and gain scheduling was drawn in \cite{moradi2007control}, where feedback linearisation in combination with a PI controller showed significant improvement in terms of performance and robustness in the control of a nonlinear boiler-turbine unit. In the authors' view, the promise of feedback linearisation based on Lie algebra has never been solidly realised since its inception in the early eighties. We explain this by the fact that the underlying methodology assumes the availability of a reliable nonlinear first-principle-based model, that is resource-consuming to develop and bound to a limited accuracy. {Estimating models from experimental data offers a solution, but when reviewing literature we see that classical feedback linearisation \cite{sastry,khalil,wagg} never assumes data-driven models. We hypothesise that this is due to the fact that the Lie algebra techniques applied to data-driven models often turn out to be practically infeasible. {A more detailed analysis on issues related to classical feedback linearisation for data-driven models is provided in Section \ref{appendix:FBL}.}\\

% To overcome the issues related to first-principle modelling, we proceed by learning our models from data. Where physical models fail, these so-called data-driven models have the potential to accurately capture the ever-increasing nonlinearities in complex engineering systems. For this reason, system identification of nonlinear mechanical systems has gained significant popularity over the past two decades \cite{kerschen2006past,noel2017nonlinear}. When reviewing literature however, we see that classical feedback linearisation \cite{sastry,khalil,wagg} never assumes data-driven models. We hypothesise that this is due to the fact that the Lie algebra techniques applied to data-driven models often turn out to be practically infeasible. {A more detailed analysis on issues related to classical feedback linearisation for data-driven models is provided in Section \ref{appendix:FBL}.}\\ 

In this paper, we propose a novel and effective approach that allows for feedback linearisation exploiting data-driven models. The linearising controller uses a model-based reference tracking framework, designed such that it is robust to model uncertainty and imperfect state information. {We stress that our approach merely linearises input-output dynamics of nonlinear systems, it does not act as the outer-loop controller. Just like classical feedback linearisation, it is intended to let engineers apply their favourite linear control solution to a problem that was originally nonlinear.} {The specific focus of this work is on mechanical systems, which often exhibit nonlinear behaviours including harmonics and frequency-energy dependence. These phenomena can only be captured by a model class that contains a feedback path \cite{schoukens2017identification}. It is for this reasons that nonlinear mechanical systems are often modelled as LTI systems with a negative feedback path containing static nonlinearities, see, e.g., \cite{noel3,shakib2022computationally,schoukens2018linear}. A block scheme representation of this model class is illustrated in Fig. \ref{fig:underlying}.} Consider, for instance, the mass-spring-damper system with a cubic spring nonlinearity
\begin{equation}
    m\Ddot{y}(t)+c_l\Dot{y}(t)+k_ly(t)+k_cy^3(t)=u(t),
    \label{eq:duf}
\end{equation}
where the output $y(t)$ is the displacement of the mass $m$, and the input $u(t)$ is a force exiting the mass. Moreover, $c_l$ and $k_l$ are the linear damping and stiffness constants, respectively; $k_c$ is the cubic spring coefficient and the over-dot indicates a derivative with respect to the time variable $t$. The dynamics can easily be rewritten to obey the structure of Fig. \ref{fig:underlying}, as
\begin{equation}
    m\Ddot{y}(t)+{c_l}\Dot{y}(t)+k_l y(t)=u(t)-k_c y^3(t),
    \label{eq:exsystem}
\end{equation}
suggesting that the cubic spring nonlinearity acts as an additional input to the underlying linear system. A system that can be modelled in this manner allows for a more careful selection of the nonlinear basis functions for system identification, since it is known that the nonlinearities are functions of the outputs only. Furthermore, these models clearly distinguish the underlying linear dynamics in the feedforward path from the undesirable nonlinearities in the feedback branch. In the case of vibrating systems, it is known that the linear dynamics are stable and contain the important vibration properties of the system, namely its linear modal properties. In this study, we treat systems can be represented by the model structure of Fig. \ref{fig:underlying}.\\

\begin{figure}[t]
    \centering
    \includegraphics{ 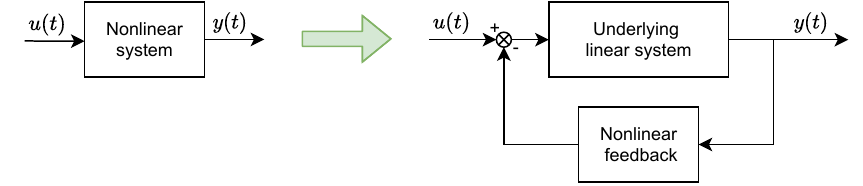}
    \caption{Illustration of the model structure adopted in this work. A nonlinear mechanical system is modelled as an underlying linear system with a static nonlinear output function wrapped around it in feedback.}
    \label{fig:underlying}
\end{figure}

In summary, the three main contributions of the paper are:
\begin{itemize}
    \item the development of a new framework that exploits data-driven models and robust reference tracking towards linearising the IO response of nonlinear {feedback systems with output nonlinearities}.
    \item the preservation in this framework of the underlying LTI response of the system, or alternatively, of its linear vibration properties.
    \item the quantification of the linearising performance over a wide range of frequencies by means of a nonparametric nonlinear distortion analysis.
\end{itemize}

The paper is organised as follows: firstly, {a more detailed introduction to classical feedback linearisation and its limitations for data-driven models is provided in Section \ref{appendix:FBL}.} Then, Section \ref{sec:problem} presents the proposed approach, including both the control design method as well as the data-driven modelling and analysis tools. Section \ref{sec:simulations} studies an extensive set of simulation examples in order to demonstrate the performance and robustness properties of the introduced framework. In Section \ref{sec:experiments}, the complete control structure is validated by means of experimental tests on a prototype of a high-precision motion system. Finally, conclusions are drawn and recommendations for future works are discussed in Section \ref{sec:conclusion}.
    \section{Classical feedback linearisation and its limitations} \label{appendix:FBL}
Classical feedback linearisation \cite{khalil} is well-defined for SISO systems that take the following state-space form:
    \begin{align}
        \dot{{x}}&=\mathbf{f}({x})+\mathbf{g}({x}) {u}, \nonumber\\
        {y}&=\mathbf{h}({x}),
        \label{eq:fblin}
    \end{align}
    where $\bm{f}(\cdot)$, $\bm{g}(\cdot)$ and $\bm{h}(\cdot)$ are (non)linear functions of the state vector $x$.
    The underlying procedure of feedback linearisation relies on repeatedly differentiating the output $y$ until the input $u$ explicitly appears. Differentiating the output with respect to time yields
    \begin{equation}
        \dot{{y}}=\frac{\partial \mathbf{h}({x})}{\partial {x}} \dot{{x}} = \frac{\partial \mathbf{h}({x})}{\partial {x}} \mathbf{f}({x})+\frac{\partial \mathbf{h}({x})}{\partial {x}} \mathbf{g}({x}){u}=L_{{f}}\mathbf{h}({x})+L_{{g}}\mathbf{h}({x}){u},
        \label{eq:lie1}
    \end{equation}
    where $L_{{f}}\mathbf{h}({x})$ and $L_{{g}}\mathbf{h}({x})$ are the so-called Lie derivatives of $\mathbf{h}({x})$ with respect to $\mathbf{f}({x})$ and $\mathbf{g}({x})$. The feedback linearising input to the nonlinear system is then defined as
    \begin{equation}
        {u}^*=\frac{1}{L_{g} L_{f}^{\rho-1} \mathbf{h}({x})}(-L_{f}^{\rho} \mathbf{h}({x})+{v}),
        \label{eq:ustar}
    \end{equation}
    in which the linear relation between output and outer-loop control input ${v}$ becomes ${y}^{\rho}={v}$, where the notation ${y}^{\rho}$ indicates the $\rho$th derivative of ${y}$. Here, $\rho$ denotes the relative degree, i.e. the number of times the output had to be differentiated until the input appeared. Thus, by means of feedback linearisation, the nonlinear system is converted to a linear equivalent, which is a chain of $\rho$ integrators.\\

    As an example, consider the analytical state-space model of the nonlinear mass-spring-damper system (\ref{eq:duf}), which obeys (\ref{eq:fblin}) with $x = \left[y\;\; \dot{y}\right]^{\top}$:
    \begin{equation}
    \begin{cases}
        \begin{aligned}
            \dot{x}&=\left[\begin{array}{cc} 0 & 1\\-k_l/m & -c_l/m\end{array}\right] x
            +\left[\begin{array}{c} 0\\ 1/m\end{array}\right]u+
            \left[\begin{array}{c}  0\\-k_c/m\end{array}\right] x_1^3\\
            y&=\left[\begin{array}{cc} 1 & 0\end{array}\right] x,
            \label{eq:nlplant}
        \end{aligned}
        \end{cases}
    \end{equation}
     To find the linearising control law, we first differentiate the output until the input appears:
    \begin{subequations}
    \begin{align}
        y &= x_1,\\
        \dot{y} & = \dot{x}_1 = x_2,\\
        \Ddot{y} & = \dot{x}_2 = - \frac{k_l}{m} x_1 - \frac{c_l}{m} x_2 - \frac{k_c}{m} x_1^3 + \frac{1}{m} u.
    \end{align}\label{eq:ydd}
    \end{subequations}
    The relative degree equals 2, so we intend to find a new input that transforms the IO relation to a double integrator, i.e., $\ddot{y}=v$. Using (\ref{eq:ustar}), we find the linearising feedback control law as
    \begin{equation}
        u^* = {k_l}x_1 + {c_l} x_2 + {k_c} x_1^3 + mv.
    \end{equation}

    Feedback linearisation in the described form suffers from drawbacks that hamper its widespread usage. We proceed by listing the six main drawbacks, including some commonly known ones as well as those specific to data-driven models and nonlinear mechanical systems:
    \begin{enumerate}
        \item [{D1}] \textbf{Linear dynamics are not preserved.} Classical feedback linearisation transforms the IO relation of any system into a chain of integrators. From a physical point of view, it would however be desirable to maintain the modal properties of the system, since they contain important information. From a control perspective it would also be more intuitive to preserve those properties if one wishes to design a controller using, for example, loop-shaping techniques. 
        \item [{D2}] \textbf{Digital implementation is not considered.} The mathematical foundation of feedback linearisation is based on continuous-time models of continuous-time systems. However, in practice, the linearising control law is implemented digitally, even while their design is based upon a continuous-time model. Research has shown that the effects of sampling and holding impose severe restrictions \cite{arapostathis} and may even quickly result in the impossibility to linearise \cite{grizzle}. The classical approach does not take into account the implementation on a digital device. {Note, however, that discrete-time equivalents of feedback linearisation exist (see, e.g., \cite{monaco1987minimum,aranda1996linearization,jakubczyk1987feedback}), but only for inherently discrete-time systems, not for discretised versions of continuous-time models.}
        \item [{D3}] \textbf{Model/plant mismatch is not accounted for.} Classical feedback linearisation does not inherently account for the mismatch between model and system that arises in practice. The analytical model is given and considered to be perfect. Modelling errors and/or imperfect state estimation are therefore not adequately accounted for which results in the presence of some of the undesirable nonlinearities in the IO response. Robustness issues are then typically addressed by the outer-loop controller.  
        \item [{D4}] \textbf{Linearising framework relies on analytical models.} A cornerstone of the underlying methodology of classical feedback linearisation is the notion of the relative degree: it is directly related to how the input enters a system, and thus has a physical interpretation. This interpretation is coherent with analytical models derived from first-principles, but it is lost when considering data-driven models. In a data-driven equivalent of (\ref{eq:nlplant}), every matrix element is nonzero. If we would then follow (\ref{eq:ydd}), we see the input explicitly appearing after differentiating the output once, which implies that the relative degree is always equal to 1. 
        \item [{D5}] \textbf{Internal dynamics are potentially unstable.} In cases where the relative degree is smaller than the system order (which is always the case for data-driven models), the state equation will only be partially linearised \cite{khalil}, meaning that there may be some remaining nonlinear dynamics in the system which do not have a direct influence on the IO relation. Yet, these remaining dynamics can cause internal instability, so global stability of the internal dynamics must be guaranteed. This can be achieved by considering input-to-state stability \cite{bos}. However, this may result in additional bounds on the outer-loop input $v(t)$.
        \item [{D6}] \textbf{Control law is potentially unfeasible.} If we neglect D4 and D5 for the moment, we could still attempt to use classical feedback linearisation with a data-driven model. However, due to the underlying methodology, this would be an insensible choice since classical feedback linearisation attempts to exactly eliminate all nonlinearities, meaning that any nonlinear non-physical parameter of the data-driven model is intended to be cancelled out. Imagine, for example, that in (\ref{eq:nlplant}), $y=x_1 + \lambda x_2$, where $\lambda$ is arbitrarily small such that its effect on the output is negligible. Using (\ref{eq:ustar}), we find the linearising feedback control law that transforms the IO response into a single integrator as:
        \begin{equation}
            u^* = {k_l}x_1 + {c_l} x_2 + {k_c} x_1^3 + \frac{m(v - x_2)}{\lambda}.
            \label{eq:us}
        \end{equation}
        Although theoretically possible, this control input is practically infeasible since we divide the last term by an arbitrarily small number, thus rendering an excessively high input. Whereas this example is already problematic, such phenomena prevail even more when the model order grows. In the present form, there is no guarantee that physically reachable control signals will be generated.
    \end{enumerate}
    The linearisation framework of this paper attempts to address the above drawbacks by proposing a novel viewpoint towards feedback linearisation of nonlinear mechanical systems.
    \color{black}
    
    \section{Data-driven feedback linearisation framework} \label{sec:problem}
This section introduces the methodologies that span the proposed linearisation framework. First, the control strategy is described in detail. Next, a brief overview is given regarding the data-driven modelling and analysis tools.

\subsection{Control design} \label{sec:control}
The dynamic models considered in this work are purely data-driven. More specifically, the focus is on discrete-time SISO nonlinear state-space models, defined at time instant $k \in \mathbb{N}$ as 
\begin{equation}
    \begin{cases}
        \begin{aligned}
            x(k+1)&=\bm{A} x(k)+\bm{B} u(k)+\bm{E} \zeta(y(k))\\
            y(k)&=\bm{C} x(k),
            \label{eq:nleq}
        \end{aligned}
    \end{cases}
\end{equation}
where $\bm{A} \in \mathbb{R}^{n \times n}$, $\bm{B} \in \mathbb{R}^{n \times 1}$ and $\bm{C} \in \mathbb{R}^{1 \times n}$ are the linear state, input, and output matrices, respectively, and with $n$ being the model order. Furthermore, $x(k) \in \mathbb{R}^n$ is the state vector, $u(k) \in \mathbb{R}$ the plant input and $y(k) \in \mathbb{R}$ the plant output; $\zeta(y(k)) \in \mathbb{R}^s$ is the undesired nonlinear function of the output, with $\bm{E} \in \mathbb{R}^{n \times s}$ the associated coefficient matrix. The model in the form of (\ref{eq:nleq}) corresponds to the structure schematised in Fig. \ref{fig:underlying}, with $\bm{E}\zeta(y(k))$ being the undesired static nonlinearity in the feedback loop.\\

{This nonlinearity is to be eliminated from the IO response, while the underlying linear dynamics are preserved. To this end, an internal reference tracking framework is proposed, where the inner-loop reference signal depends on the outer-loop input $v(k)$. For stability and performance reasons, it is typically preferred for the inner-loop to run at a higher sample rate than the outer-loop. In the proposed framework, the inner-loop runs at a sample rate which is an integer multiple greater than the outer-loop sample rate. The specific workflow is that the outer-loop sample time $t_{s,out}$ is considered given or is chosen first such that it meets performance requirements of the desired closed-loop bandwidth. Next, $t_{s,in}$ can be selected accordingly, considering the relation $N_{p,\max}=t_{s,out}/t_{s,in} \in \mathbb{N}$. Typically, when $t_{s,in}$ decreases, reference tracking and disturbance rejection improve up until a certain height where they plateau. This plateauing point is usually adopted for $t_{s,in}$, assuming that the computational effort is reasonable. Then, inner-loop references are generated as follows:}
\begin{equation}
    \begin{cases}
        \begin{aligned}
            x_{ref}(i+1|k)&=\bm{A} x_{ref}(i|k)+\bm{B} v(k)\\
            y_{ref}(i+1|k)&=\bm{C} x_{ref}(i+1|k),
            \label{eq:yref}
        \end{aligned}
    \end{cases}
\end{equation}
where $x_{ref}(i+1|k)$ and $y_{ref}(i+1|k)$ are the future state and output sub-reference points based on the outer-loop input $v(k)$, for $i\in\{0,\ldots,N_{p,\max}-1\}$. Moreover, note that $\bm{A}$ and $\bm{B}$  here correspond to sample time $t_{s,in}$, {and that $(i|k)$ indicates a unique instance in time that combines the inner and outer-loop sample times (e.g., the current time instance in Fig. \ref{fig:dynamicN} is $(2|k)$)}. A consequence of the proposed solution is that the prediction horizon has a dynamic length, because the outer-loop input $v(k)$ is supplied at a lower rate, as is illustrated in Fig. \ref{fig:dynamicN}.

A controller that is capable of perfectly tracking (\ref{eq:yref}) at any time $(i|k)$ for any input $v(k)$, generates an IO relation between $v(k)$ and $y(k)$ that is governed by the linear part of the identified model. {This addresses drawback D1 of classical feedback linearisation while it is still in line with its rationale of Fig. \ref{fig:FBLscheme}.} {Note that although it is sensible to exactly preserve the linear vibrating properties of the system under test, the reference signal is not restricted to take the specific form of  (\ref{eq:yref}). In theory, any custom reference signal can be chosen, which increases the flexibility of the proposed method.}\\

% This nonlinearity is to be eliminated from the IO response, while the underlying linear dynamics are preserved. To this end, an internal reference tracking framework is proposed that follows a one-step-ahead reference for any outer-loop input $v(k)$. This reference point, which is generated online at any time instant $k$, is the one-step-ahead output response of the desirable linear part of the model to $v(k)$. In mathematical terms, the one-step-ahead reference point at time instant $k$ is defined as follows:
% \begin{equation}
%     \begin{cases}
%         \begin{aligned}
%             x_{ref}(k+1)&=\bm{A} x_{ref}(k)+\bm{B} v(k)\\
%             y_{ref}(k+1)&=\bm{C} x_{ref}(k+1),
%             \label{eq:yrefo}
%         \end{aligned}
%     \end{cases}
% \end{equation}
% where $x_{ref}(k+1)$ is the state reference and $y_{ref}(k+1)$ is the output reference to be tracked. So, a controller that is capable of perfectly tracking this one-step-ahead reference point at any time $k$ for any input $v(k)$, generates an IO relation between $v(k)$ and $y(k)$ that is governed by the linear part of the identified model. {This addresses drawback D1 of classical feedback linearisation while it is still in line with its rationale of Fig. \ref{fig:FBLscheme}.} \color{red}{Note that although it is sensible to preserve the linear vibrating properties of the system under test by tracking (\ref{eq:yrefo}), the reference points are not restricted to this specific form. In theory, any custom reference dynamics can be chosen, which increases .} \color{black}\\

\begin{figure}[!t]
    \centering
    \includegraphics[width=.6\textwidth]{  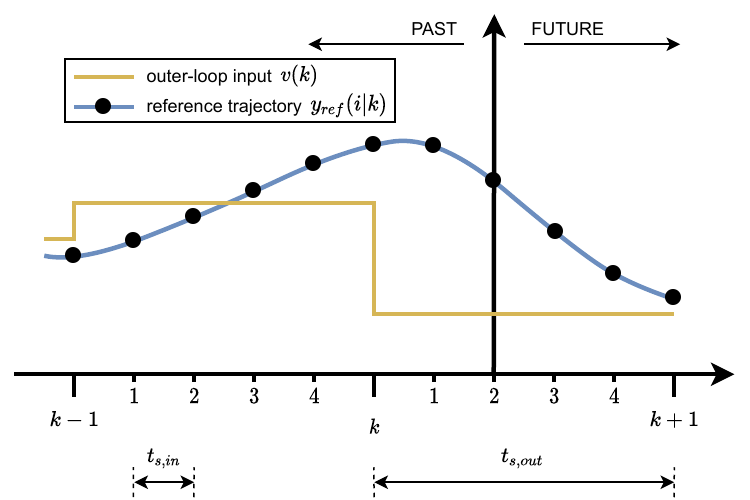}
    \caption{Illustration of the proposed solution with online reference generation and dynamic horizon length. The notation $y_{ref}(i|k)$ indicates the output reference point at the $i$th inner-loop sub-sample of the $k$th outer-loop time instant. At the current time instant, there are 3 out of 5 future references available. The number of available references reduces to 1 as time progresses up until time $k+1$, where a new outer-loop input will generate 5 new reference points.}
    \label{fig:dynamicN}
\end{figure}

{We propose a linearising controller inspired by unconstrained model predictive control \cite{rossiter2003model,camacho2013model,garcia1989model,allgower2012nonlinear}. Moreover, {to overcome drawback D3,} the controller has integral action in the sense that robust convergence can be achieved in case of modelling errors and imperfect state information. To obtain this, the original model is augmented such that the state equation also contains the current measured output, as inspired by the velocity form approach in \cite{gonzalez2008conditions,wang2004tutorial,betti2013robust}.}
Consider again the $n$th order SISO discrete-time nonlinear model
\begin{equation}
    \begin{cases}
        \begin{aligned}
            x(i+1|k)&=\bm{A} x(i|k)+\bm{B} u(i|k)+\bm{E}\zeta(y(i|k))\\
            y(i|k)&=\bm{C} x(i|k),
            \label{eq:greybox2}
        \end{aligned}
    \end{cases}
\end{equation}
corresponding to sample time $t_{s,in}$. In order to augment the state space, the following relations are defined:
\begin{equation}
    \begin{split}
        \begin{cases}
        \Delta x(i|k) &:= x(i|k) - x(i-1|k)\\
        \Delta u(i|k) &:= u(i|k) - u(i-1|k)\\
        \Delta \zeta(y(i|k)) &:= \zeta(y(i|k)) - \zeta(y(i-1|k)).
        \end{cases}
    \end{split}
\end{equation}
Next, the state-space model (\ref{eq:greybox2}) in augmented form writes
\begin{equation}
    \begin{cases}
        \begin{aligned}
            \olsi{x}(i+1|k)&=
            \underbrace{\left[\begin{array}{cc}
            \bm{A} & \bm{0}_{n \times 1} \\
            \bm{C} \bm{A} & 1
            \end{array}\right]}_{\olsi{\bm{A}}}\olsi{x}(i|k)
            +
            \underbrace{\left[\begin{array}{c}\bm{B} \\
            \bm{C} \bm{B}
            \end{array}\right]}_{\olsi{\bm{B}}} \Delta u(i|k)
            +
            \underbrace{\left[\begin{array}{c}\bm{E} \\
            \bm{C} \bm{E}
            \end{array}\right]}_{\olsi{\bm{E}}}\Delta \zeta(y(i|k))\\
            y(i|k)&=
            \underbrace{\left[\begin{array}{cc}
            \bm{0}_{1 \times n} & 1
            \end{array}\right]}_{\olsi{\bm{C}}}\olsi{x}(i|k),
        \end{aligned}
    \end{cases}
    \label{eq:augmstatespace}
\end{equation}
where $\olsi{x}(i|k)=\left[\Delta x(i|k)^{\top} \:\: y(i|k)\right]^{\top}$.
At time $t(i|k)$, the predicted future outputs of (\ref{eq:augmstatespace}) over prediction horizon $N_p$, \textit{i.e.}, 
\begin{equation}
    Y_k = \left[y(i+1|k), \ldots, y(i+N_p| k)\right]^{\top} \in \mathbb{R}^{N_p},
\end{equation}
is a function of the current state, input sequence and disturbance sequence; it is defined as
\begin{equation}
    Y_k = S_x\olsi{x}(i|k) + S_u\Delta U_k + S_g \Delta G_k,
    \label{eq:predict}
\end{equation}
where $S_x$, $S_u$ and $S_g$ are auxiliary matrices that describe the evolution of the states and output over the prediction horizon. A detailed derivation of these matrices is provided in \ref{appendix:MPC}. Furthermore,
\begin{equation}
    \Delta U_k = \left[\begin{array}{c}
    \Delta u(i|k)\\
    \vdots\\
    \Delta u(i+N_p-1|k)\\
    \end{array}\right]\in \mathbb{R}^{N_p};
    \quad
    \Delta G_k = \left[\begin{array}{c}
    \Delta \zeta(y(i|k))\\
    \vdots\\
    \Delta \zeta(y(i+N_p-1|k))\\
    \end{array}\right]\in \mathbb{R}^{sN_p},
\end{equation}
where $\Delta U_k$ is the vector to be optimised,
% and $\Delta G_k$ is the vector of future disturbances (the word disturbance is used because the elements of $\Delta G_k$ cannot be controlled). There is a conflicting situation between $\Delta G_k$ and $Y_k$ in the sense that they are mutually dependent. Fortunately, $\Delta G_k$ can very well be estimated in advance, by substituting the available reference information of (\ref{eq:yref}) for the future outputs. Thus, the estimate of $\Delta G_k$ becomes
% \begin{equation}
%    \Delta \widetilde{G}_k = \left[\begin{array}{c}
%     \zeta(y(i|k))-\zeta(y(i-1|k))\\
%     \zeta(y_{ref}(i+1|k))-\zeta(y(i|k))\\
%     \vdots\\
%     \zeta(y_{ref}(i+N_p-1|k))-\zeta(y_{ref}(i+N_p-2|k))
%     \end{array}\right].
% \end{equation}
% By doing so, it is assured that $Y_k$ can be computed solely based on available data and the to be optimised input sequence $\Delta U_k$.
for which the following quadratic cost function is proposed:
\begin{equation}
    J = (Y_k - Y_{k,ref})^{\top}\Omega (Y_k - Y_{k,ref})
        + \Delta U_k^{\top} \Psi \Delta U_k,
    \label{eq:cost}
\end{equation}
where $Y_{k,ref}$ is obtained from (\ref{eq:yref}), and
\begin{equation}
    \Omega = \left[\begin{array}{cccc}
    Q & 0 & \cdots & 0\\
    0 & Q & \cdots &0\\
    \vdots & \vdots & \ddots & \vdots\\
    0& 0  & \cdots & Q     
    \end{array}\right] \in \mathbb{R}^{N_p \times N_p};
    \quad
    \Psi = \left[\begin{array}{cccc}
    R_{\Delta} & 0 & \cdots & 0\\
    0 & R_{\Delta} & \cdots &0\\
    \vdots & \vdots & \ddots & \vdots\\
    0& 0  & \cdots & R_{\Delta}     
    \end{array}\right] \in \mathbb{R}^{N_p \times N_p},
\end{equation}
with $Q$ being a positive scalar penalising any deviation from the reference and $R_{\Delta}$ being a positive scalar penalising the rate of change of the plant input. These values are typically chosen following performance specifications. {In essence, the values of $Q$ and $R_{\Delta}$ can be tuned such that a satisfactory balance between tracking accuracy and controller aggressiveness is obtained.}
% Moreover, from (\ref{eq:yref}),
% \begin{equation}
%     Y_{k,ref} = \left[y_{ref}(i+1|k), \ldots, y_{ref}(i+N_p| k)\right]^{\top} \in \mathbb{R}^{N_p}.
% \end{equation}

{Minimising (\ref{eq:cost}) is a nonlinear optimisation problem that needs to be solved online, at every time step, which is not in line with the rationale of feedback linearisation that aims to simplify control design. Luckily, the proposed model class with output nonlinearities allows us to exploit future reference information to approximate $\Delta G_k$ as 
\begin{equation}
   \Delta \widetilde{G}_k = \left[\begin{array}{c}
    \zeta(y(i|k))-\zeta(y(i-1|k))\\
    \zeta(y_{ref}(i+1|k))-\zeta(y(i|k))\\
    \vdots\\
    \zeta(y_{ref}(i+N_p-1|k))-\zeta(y_{ref}(i+N_p-2|k))
    \end{array}\right].
\end{equation}
If we then replace $\Delta G_k$ by $\Delta \widetilde{G}_k$ and substitute (\ref{eq:predict}) in (\ref{eq:cost}), the cost function $J$ becomes a quadratic and convex function of $\Delta U_k$ {with a closed-form solution.} Hence, by setting the derivative of $J$ with respect to $\Delta U_k$ to zero and solving for $\Delta U_k$, the optimal input sequence that leads to the global minimum of the cost function is found to be
\begin{subequations}
    \begin{align}
    \Delta U^*_k &= -W^{-1} F\big(S_x\olsi{x}(i|k) + S_g \Delta \widetilde{G}_k - Y_{k,ref}\big),
    \label{eq:optu}
\intertext{where}
        \begin{split}
            W&=2\big(\Psi + S_u^{\top} \Omega S_u\big),
        \end{split}\\
        \begin{split}
            F&= 2S_u^{\top}\Omega.
        \end{split}
    \end{align}\label{eq:optu_}
\end{subequations}
Finally, the control input that is applied to the plant at time $t(i|k)$ is defined as
\begin{equation}
    u(i|k) = u(i-1|k) + \left[1 \;\; \bm{0}_{1 \times N_p-1} \right] \Delta U^*_k. \label{eq:inputt}
\end{equation}
{Note that by approximating $\Delta G_k$, the seemingly complicated nonlinear control problem simply boils down to plugging information about the states and output reference into a control law that can be pre-computed offline. Any inaccuracies that are the result of this approximation are compensated for the robust integral action property of the controller. We emphasise that the approximation of $\Delta G_k$ is only possible because of the considered model class: if there would be state or input nonlinearities it is not directly applicable. {Finally, note that the control design is inherently in discrete time, therefore overcoming drawback D2 of classical feedback linearisation.}} 

In the present form, one may argue that instability might occur when the prediction horizon approaches 1. If this is indeed the case, a minimum horizon length $>$ 1 can be assigned, which means that future references need to be estimated. A fair compromise between stability and accuracy is to assume that the last available reference point also holds for the following unknown reference points. At every new outer-loop time instant $k$, these incorrect reference points are replaced by the correct ones. The proposed method does not guarantee stability, but we argue that by appropriately tuning $Q$ and $R_{\Delta}$, and optionally assigning a minimum horizon length, the risk of instability can be mitigated to a great extent.
% Note that all auxiliary matrices that describe the evolution of the states are calculated only once, offline, for $N_p=N_{p,\max}$. Subsequently, online, the correctly sized matrices corresponding to the current value of $N_p$ are extracted from the $N_p=N_{p,\max}$ matrices. By doing so, the seemingly complicated nonlinear MPC controller simply boils down to plugging information about the states and output reference into a pre-calculated control law. In this way, the algorithm remains computationally attractive, while the inegral action property. 

\begin{figure}[!b]
    \centering
    \includegraphics[width=\textwidth]{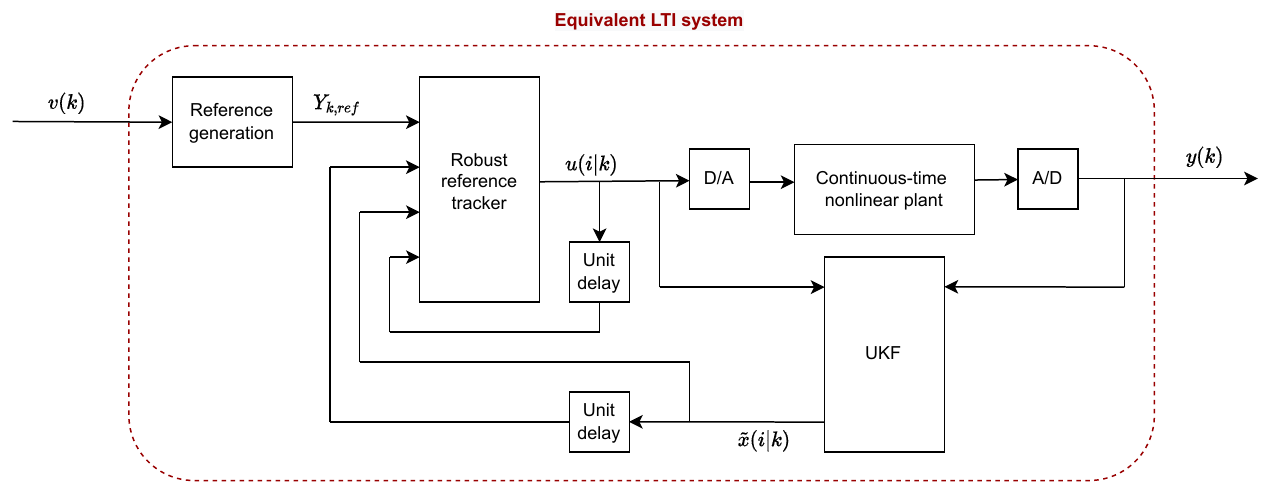}
    \caption{Schematic overview of internal reference tracking framework that generates a linearised IO map between $v(k)$ and $y(k)$.}
    \label{fig:mpcscheme}
\end{figure}

From (\ref{eq:optu}) can be seen that the proposed solution needs the current (and previous) state information in order to calculate the input signal. Whereas only the output is measured, a nonlinear observer is needed to estimate the states. In this work, state estimation is performed by means of the unscented Kalman filter (UKF), about which the reader is referred to \cite{julier1995new,julier1997new,wan2000unscented,laviola2003comparison} for more information. Note, however, that the UKF introduces the hyperparameters $Q_{UKF}$ and $R_{UKF}$, which represent the process noise covariance matrix and the output noise variance, respectively. {A high-level schematic overview of the control scheme is depicted in Fig. \ref{fig:mpcscheme}, while Algorithm \ref{alg} provides a lower-level summary of the proposed multi-rate control method.}\\

{Note that in the proposed method, drawbacks D4-D6 are not explicitly addressed. However, due to the inherent nature of our method when compared to the classical approach (reference tracking vs. exact cancellation), D4 and D5 are automatically accounted for. As for D6, like any model-based control method that uses data-driven models, we argue that the feasibility of the control law is insensitive to the exact values of the matrix elements. Moreover, since we preserve the linear dynamics, the controller only has to correct for the undesired nonlinearities. We therefore hypothesise that the linearisation does not require excessive inputs. We emphasise again that the aggressiveness of the controller can be tuned by modifying the cost matrices $Q$ and $R_\Delta$ in
order to remain in feasible actuator regions.}

{
\begin{algorithm}
\caption{Multi-Rate Control Scheme}\label{alg}
\begin{algorithmic}[1]{
    \State \textbf{Input:} Nonlinear state-space model of the form (\ref{eq:nleq})
    \State Define inner-loop sampling time $t_{s,in}$ based on outer-loop sampling time $t_{s,out}$ 
    \State Select hyperparameters $Q$, $R_{\Delta}$, $Q_{UKF}$, and $R_{UKF}$
    \For{$k=0,1,2,\ldots$}
        \State Compute reference signals $\bar{Y}_{k,ref}$ based on outer-loop input $v(k)$ according to (\ref{eq:yref})
        \For{$i=0,1,\ldots, N_{p,\max}$ \textbf{where} $N_{p,\max}=t_{s,out}/t_{s,in}\in \mathbb{N}$}
            \State Estimate the current state $\bar{x}(i|k)$
            \State Set the horizon length $N_p=N_{p,\max}-i$ 
            \State Extract $Y_{k,ref} \subseteq \bar{Y}_{k,ref}$ based on $N_p$
            \State Substitute $\bar{x}(i|k)$ and $Y_{k,ref}$ in (\ref{eq:optu_}) to compute $\Delta U_k^*$
            \State Apply $\Delta U_k^*$ to the system according to (\ref{eq:inputt})
        \EndFor
    \EndFor}
\end{algorithmic}
\end{algorithm}
}\color{black}

    \section{Numerical demonstration using a Duffing oscillator} \label{sec:simulations}
\begin{table}[!b]
\begin{center}
    \begin{tabular}{c||c|c|c|c|c}
         & $m$ & $c_l$ & $k_l$ & $k_q$ & $k_c$ \\
         \hline \hline
        value & 1 & 1 & $5\cdot10^2$ & $5\cdot10^4$ & $1\cdot10^8$\\
        \hline
        unit & kg & Ns/m & N/m & N/m$^2$ & N/m$^3$
    \end{tabular}
\end{center}
\caption{Physical parameters of the nonlinear plant. With these values, the natural frequency and the damping ratio of the underlying linear system are 3.56 Hz and 2.24$\%$, respectively.}
\label{table:param}
\end{table}
In this section, simulations are performed on a Duffing oscillator with a quadratic and cubic spring nonlinearity. The plant dynamics are governed by 
\begin{equation}
    m\Ddot{y}(t)+c_l\Dot{y}(t)+k_l y(t)+k_q y^2(t)+k_c y^3(t)=u(t),
    \label{eq:plant}
\end{equation}
where $m$ is the mass, $c_l$ is the linear damping coefficient and $k_l$, $k_q$ and $k_c$ the linear, quadratic and cubic stiffness coefficients, respectively. The output $y(t)$ is the displacement of the mass, and the input $u(t)$ is a force exciting the mass. The system parameters are listed in Table \ref{table:param}.\\

\subsection{Performance quantification} \label{chap:nldid}
A nonparametric nonlinearity analysis is performed in order to validate the effectiveness of the proposed framework over the entire specified frequency range of operation, effectively quantifying the performance of the linearisation. The analysis is entirely carried out in the frequency domain, based on the measurement of frequency response functions (FRFs). In this work, we consider excitation signals in the form of random-phase multisines \cite{schoukens2019nonlinear}. They are defined as
\begin{equation}
    {u}(t)=\frac{1}{\sqrt{N}} \sum_{q=-N / 2+1}^{N / 2-1} U_{q} e^{j\left(2 \pi q f_{res} t+\varphi_{q}\right)},
    \label{eq:rpms}
\end{equation}
with $\varphi_{-q}=-\varphi_q$, $U_{-q}=U_q$, $U_0=0$ and with a frequency resolution $f_{res}=f_s/N=1/T$, where $f_s$ is the sampling frequency and $T$ the period of the multisine. The phases $\varphi_q$ are independent and uniformly distributed on the interval $[0,2 \pi)$, such that $E\{\varphi_q\}=0$. Moreover, $U_q$ is the excitation amplitude, $j$ is the imaginary unit and $N$ is the number of time samples, while the division by $\sqrt{N}$ normalises the root-mean-square (RMS) value of the excitation signal. Random-phase multisines are partly deterministic in the frequency domain where their amplitude spectrum can be freely adjusted, but appear like random signals in the time domain.\\

\begin{figure}[!b]
    \centering
    \includegraphics[width=0.9\textwidth]{  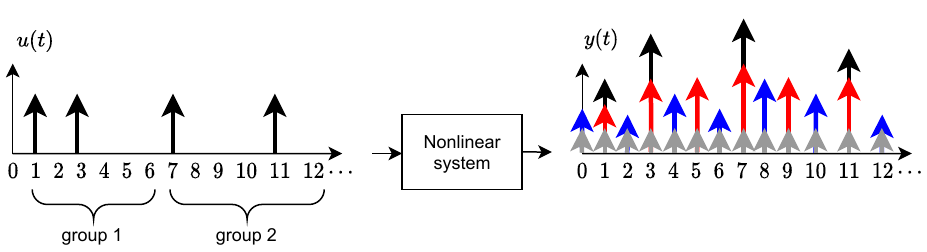}
    \caption{Illustration of the nonlinear distortion analysis. Left: input spectrum of the multisine. Right: the output spectrum with the quantification of the odd (in red) and even (in blue) nonlinear distortions. The grey lines represent the contributions of the disturbing noise and black lines are the linear system response.}
    \label{fig:nonlindist}
\end{figure}

When it comes to assessing the linearisation performance, multisines are modified such that only odd frequency lines, corresponding to odd values of $q$, are excited. Moreover, for every group of $n_f$ frequency lines, one line is randomly excluded. This rejected line will serve as an odd detection line measuring odd nonlinear distortions. All the even frequency lines serve as even detection lines, measuring even nonlinear distortions. {Odd nonlinearities are symmetric w.r.t. the origin, i.e., $-f(x)=f(-x)$ (e.g., a cubic function), while even nonlinearities are symmetric w.r.t. the y-axis, i.e., $f(x) = f(-x)$ (e.g., a quadratic function).} An illustration of the nonlinear distortion analysis is presented in Fig. \ref{fig:nonlindist}. {In this procedure, the noise spectrum is defined as the standard deviation of the FRF, averaged over the $R$ different realisations of $P$ steady-state multisine measurements.} In order to reduce the effect of noise, a sufficiently high number of realisations and steady-state periods must be processed. For more details on the nonlinearity analysis, the reader is referred to \cite{schoukens}.\\

All simulations are conducted in MATLAB/Simulink. White Gaussian noise is added to the output to simulate sensor inaccuracy with a signal-to-noise ratio (SNR) of 40 dB. From Table \ref{table:param}, it is derived that the linear natural frequency of the system is equal to 3.56 Hz. Based on this, the desired closed-loop bandwidth for an outer-loop controller is chosen equal to 14 Hz, that is well beyond the linear resonance. In other words, the frequency band of interest for the linearising controller ranges between 0 and 14 Hz. Fig. \ref{fig:nldbefore4} shows the nonlinear distortion analysis of system (\ref{eq:plant}) before linearisation. It is excited with 10 realisations of 5 periods of an odd random-phase multisine with an RMS amplitude of 0.12 N. For every group of 4 odd frequency lines, 1 line is randomly rejected to serve as an odd detection line. Moreover, 4000 points per period are considered, providing a frequency resolution of 0.025 Hz. The level of odd distortions in this figure (in red) is only 10 dB below the response level around the resonance, translating the strong activation of the nonlinear spring. Even distortions (in blue) are also seen to be non-negligible around 0 Hz, around the nonlinear resonance frequency at 3.8 Hz, and around its second harmonic at 7.6 Hz, as a result of the quadratic spring characteristic.\\

\begin{figure}
    \centering
    \includegraphics[width=.6\textwidth]{  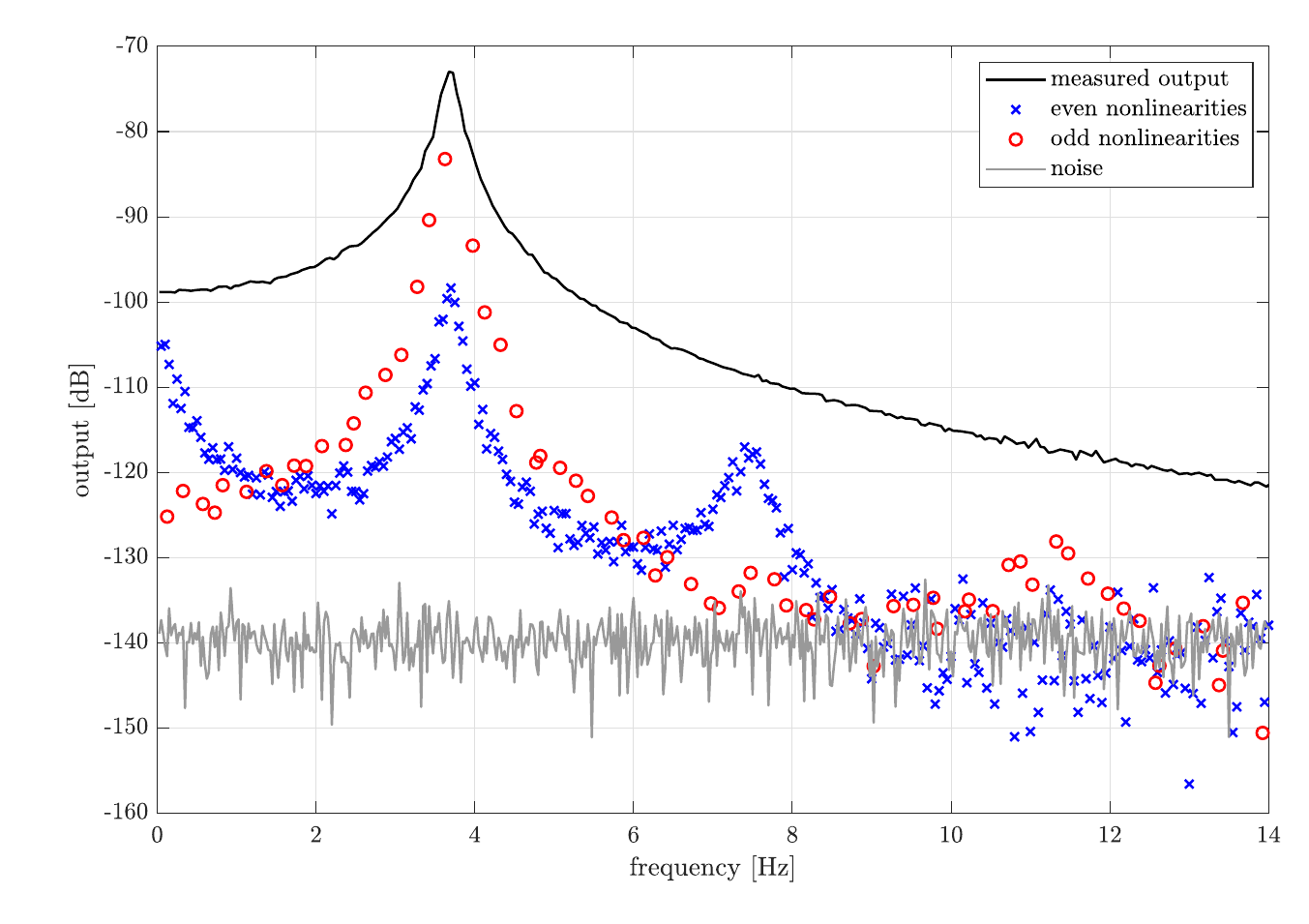}
    \caption{Nonlinear distortion analysis before linearisation.}
    \label{fig:nldbefore4}
\end{figure}

\subsection{System identification}\label{sec:nlsysid}
Even though the true system equations are known, we mimic the practical scenario of identifying models from data. Specifically, we obtain data-driven state-space models by following a state-of-the-art identification procedure consisting of 4 steps:
\begin{enumerate}
    \item \textit{Data collection:} a full multisine, \textit{i.e.}, with no detection line, is applied to the system to be linearised over multiple periods and realisations. Output data are collected and the first few periods of measurement are rejected to suppress transients.
    \item \textit{Nonparametric best linear approximation (BLA):} the nonparametric BLA is found by dividing the averaged discrete Fourier transform (DFT) of the response (output spectrum) by the averaged DFT of the excitation signal (input spectrum).
    \item \textit{Parametric BLA:} the BLA is parameterised by means of a frequency-domain subspace method \cite{mckelvey1996subspace}, weighted by the sample noise covariance matrix of the nonparametric BLA, as in \cite{pintelon2002frequency}. In this procedure, the model order is a user choice.
    \item \textit{Full nonlinear model:} the parametric BLA is used to initialise a least-squares cost function to obtain the full nonlinear model. Here, the $\bm{E}$ matrix is initially set to zero and the vector of nonlinear output monomials $\zeta(y(k))$ is constructed based on the insight obtained from the nonlinear distortion analysis and engineering judgement. 
\end{enumerate}
The identification data is obtained by performing 20 realisations of 5 periods of a full random-phase multisine with an RMS amplitude of 0.12 N. The number of data points per period is 40000 at a sampling frequency of 1000~Hz. To ensure steady-state conditions, the first period of every realisation is rejected. Moreover, 1 realisation is saved for validation. The parameter values of the obtained second-order state-space model, with $\zeta(y(i|k))=\left[y^2(i|k) \;\; y^3(i|k) \right]^{\top}$, are provided in Table \ref{table:sys1} of \ref{appendix:Models}.\\

\begin{figure}
    \centering
    \begin{subfigure}{.49\textwidth}
      \centering
      \includegraphics[width=\linewidth]{  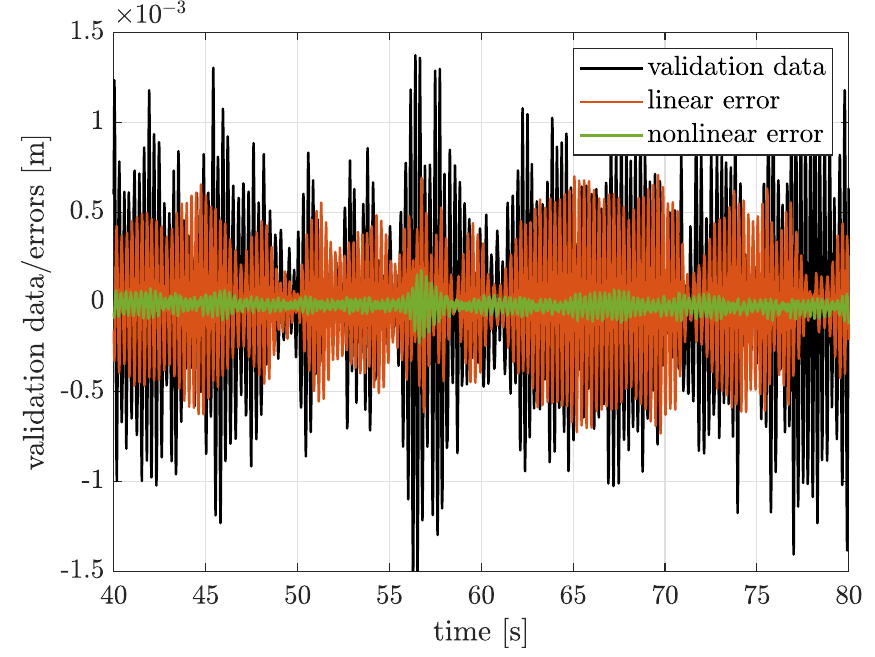}
      \caption{}
    \end{subfigure}
    \begin{subfigure}{.49\textwidth}
      \centering
      \includegraphics[width=\linewidth]{  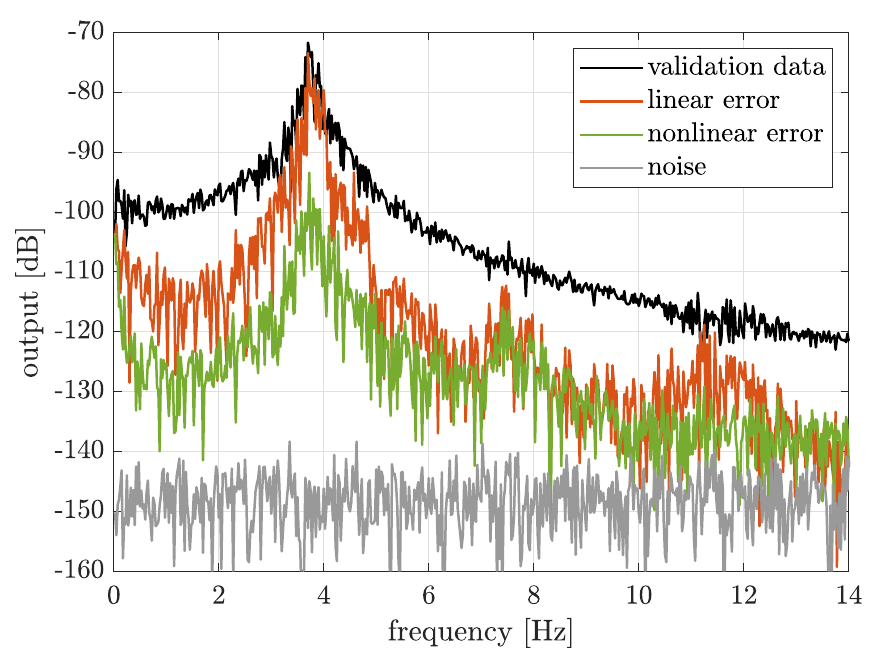}
      \caption{}
    \end{subfigure}
    \caption{(a) One period of the time-domain validation data and the associated identification errors of the parametric BLA (in orange) and of the nonlinear model (in green). (b) Corresponding frequency-domain validation plot.}
    \label{fig:sysid1m}
\end{figure}

Fig. \ref{fig:sysid1m} displays the fitting errors of the parameterised BLA (in orange) and of the full nonlinear model (in green), calculated on validation data. The left plot presents one time-domain period of the validation data with the corresponding linear and nonlinear modelling errors. The nonlinear error (6.64$\%$) appears to be significantly smaller than the linear error (59.2$\%$). These percentages are obtained by dividing the standard deviation of the errors by the standard deviation of the validation output data. The right plot in Fig. \ref{fig:sysid1m} shows the same validation data and error signals in the frequency domain. Here, the nonlinear error is consistently one order of magnitude (20 dB) smaller than the output level. The comparison between the error spectra also reveals that the nonlinear model performs well at the resonance and the third harmonic locations, while it does so less convincingly at DC and at the second harmonic location. This is explained by the smaller magnitude of the quadratic nonlinearity in the system compared to the cubic contribution.

\subsection{Linearisation performance}
As stated in Section \ref{sec:control}, the first important tuning parameter to set is the outer-loop sample time $t_{s,out}$. Since it should be at least 5 times faster than the desired closed-loop bandwidth, it is chosen as $t_{s,out}=10$ ms. The inner-loop sample time is set to $t_{s,in}=1$ ms. The values for $Q$ and $R_{\Delta}$ are chosen as $1\cdot10^{12}$ and 1, respectively. {The large difference in magnitude between the two tuning parameters is mostly due to scaling: the output displacements are small, so in order to keep the cost function (\ref{eq:cost}) well-balanced we require a high magnitude for $Q$.} As for the UKF, the measurement noise variance is set to the mean value of the output noise variance of the nonlinear distortion analysis, \textit{i.e.}, $R_{UKF}=1.13\cdot10^{-14}$ m$^2$; the process noise covariance matrix is set to $Q_{UKF}=0.05R_{UKF}\bm{I}_2$.\\

\begin{figure}[!t]
    \centering
    \includegraphics[width=.65\textwidth]{  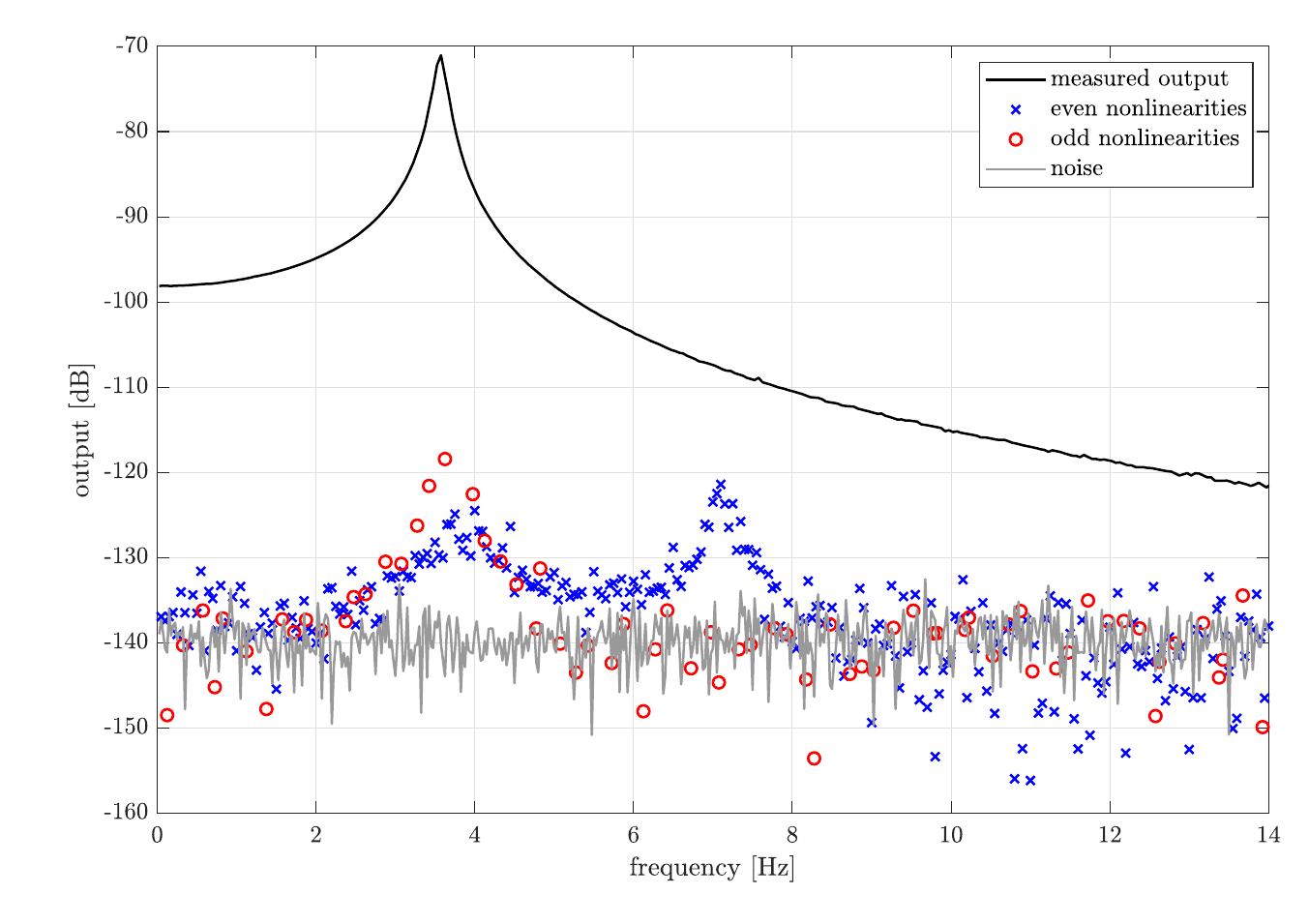}
    \caption{Nonlinear distortion analysis after linearisation. An excellent linearisation is achieved since the odd and even distortions observed Fig. \ref{fig:nldbefore4} are herein almost coincident with the noise floor.}
    \label{fig:nldafter4}
\end{figure}

Fig. \ref{fig:nldafter4} depicts the nonlinear distortion analysis of the linearised plant. The exact same multisine excitation signal as for Fig. \ref{fig:nldbefore4} is considered. An excellent linearisation is achieved since the odd and even distortions observed Fig. \ref{fig:nldbefore4} are herein almost coincident with the noise floor. The residual distortions around 3.8 and 7.6 Hz are 50 dB below the validation output. Yet, it must be noted that these  residuals can be made arbitrarily close to the noise level by tuning the controller and UKF parameters more aggressively. However, this would require excessive input energy. For the selected tuning parameters, the outer-loop input $v(i|k)$ and the plant input $u(i|k)$ are shown in Fig. \ref{fig:planti} for one period of the multisine. It can be observed from the left subplot that a reasonably large input signal is required to achieve the linearisation. In the right subplot, a zoom-in showing the underlying control actions required to achieve linearisation is provided. Seemingly, the most aggressive inputs are calculated at the start of every outer-loop sample period, \textit{i.e.}, when the prediction horizon is at its maximum value, that is $N_{p,\max}=t_{s,out}/t_{s,in} =10$.
\begin{figure}[!t]
    \centering
    \begin{subfigure}{.49\textwidth}
      \centering
      \includegraphics[width=\linewidth]{  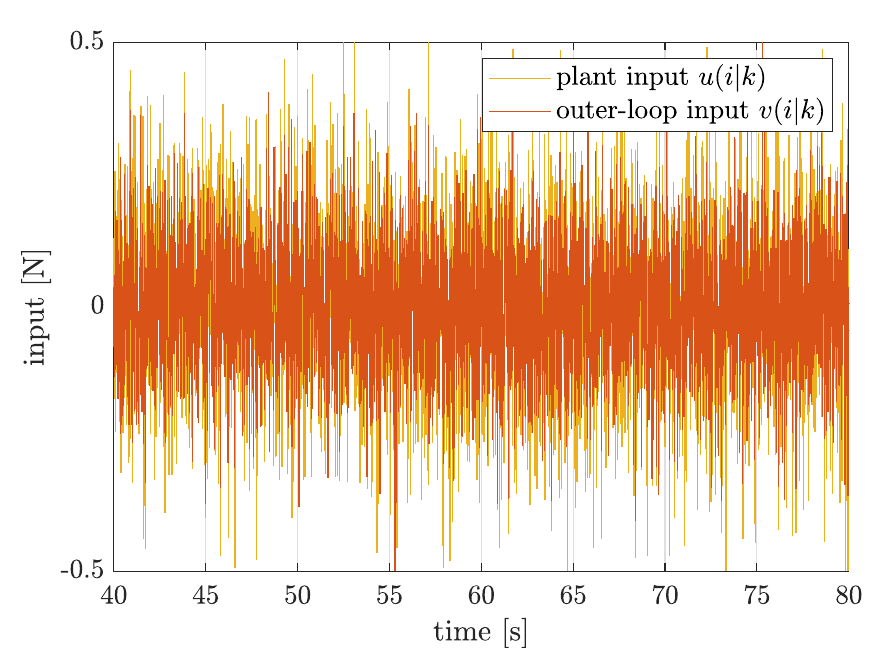}
      \caption{}
    \end{subfigure}
    \begin{subfigure}{.49\textwidth}
      \centering
      \includegraphics[width=\linewidth]{  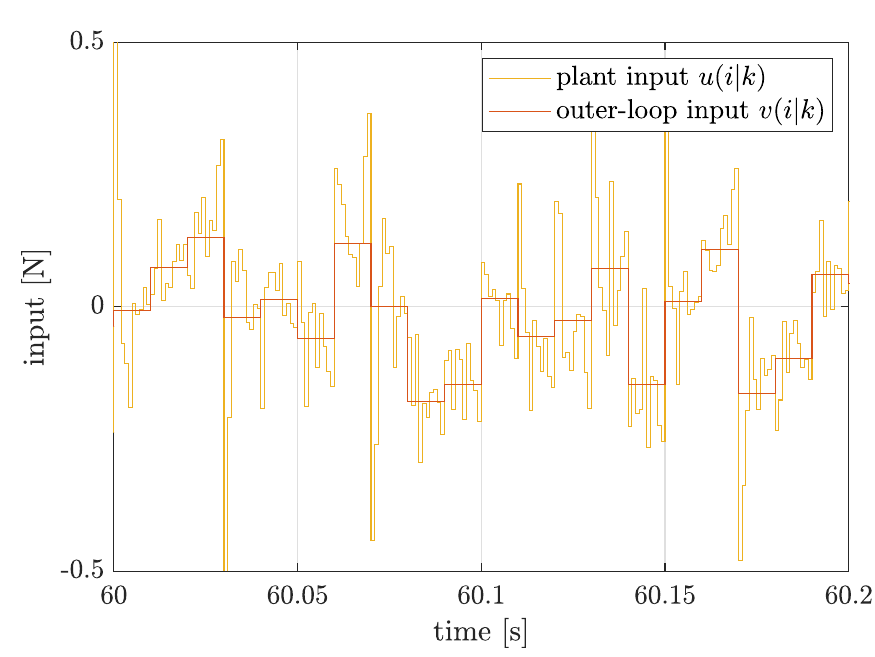}
      \caption{}
    \end{subfigure}
    \caption{(a) Time-domain inputs during one period of the random-phase multisine and (b) close-up around $t=60$ s. Both the outer-loop input (in red) and the plant input (in yellow) are shown.}
    \label{fig:planti}
\end{figure}

\subsection{Linearisation robustness}
In this section, the robustness of the proposed framework is evaluated through three distinct test. The first test provides a comparison between the proposed method and the classical approach, while the second and third test asses the linearising performance in case of unmodelled dynamics and extrapolation, respectively.

\subsubsection{Comparison to classical feedback linearisation}
Classical feedback linearisation converts the IO response into a chain of integrators, which makes a direct comparison to the proposed method cumbersome. We therefore do not follow the Lie derivatives of (\ref{eq:lie1}) and (\ref{eq:ustar}) directly, but we rather find by hand a linearising control law that follows the rationale of classical feedback linearisation while still preserving the linear dynamics. It is easy to verify that
\begin{equation}
    u(t)^* = {k_q} y(t)^2 + {k_c} y(t)^3 + v(t),
    \label{eq:fbll}
\end{equation}
provides an IO response between $v(t)$ and $y(t)$ that is governed by the desirable linear part of the system.

In order to asses the robustness of both methods, all system parameters $\theta$ in Table \ref{table:param} are subject to random perturbations. Note that we do not change the model, only the system parameters. The perturbed parameters $\hat\theta$ are determined by adding a scaled random value from a uniform distribution to the original parameter value according to:
\begin{equation}
    \hat\theta = \theta + \alpha \mathcal{U}(-\theta, \theta), 
\end{equation}
where $\alpha$ controls the severeness of the perturbation.
In Fig. \ref{fig:compare} we show the linearisation results for both methods for various values of $\alpha$. Note that we did not add synthetic measurement noise, and that we do not make the distinction between odd and even nonlinearities but show the combined level of nonlinear distortions instead. For statistical rigourness we conducted 25 distortion analyses (with the same properties as in the beginning of this section) for every value of $\alpha$.
From Fig. \ref{fig:compare} two distinct observations can be made: first, for $\alpha=0$, successful linearisation is achieved for both methods, but there are significantly more nonlinear distortions present for the classical approach. The reason for this is the digital implementation using sample and hold mechanisms (see drawback D2), which is not accounted for in the classical approach and hence results in inaccuracies. On the other hand, the proposed method directly exploits the discretised model with the right sample time, resulting in superior accuracy. A second observation is that the classical approach performs, as expected, worse for increasing values of $\alpha$ since the feedback law (\ref{eq:fbll}) does not adapt to modelling errors (see drawback D3). Contrarily, the proposed method is shown to be very robust and consistent with respect to increasing values of $\alpha$. which can be attributed to the integral action property of the linearising controller.

\begin{figure}
    \centering
    \begin{subfigure}{.49\textwidth}
      \centering
      \includegraphics[width=\linewidth]{  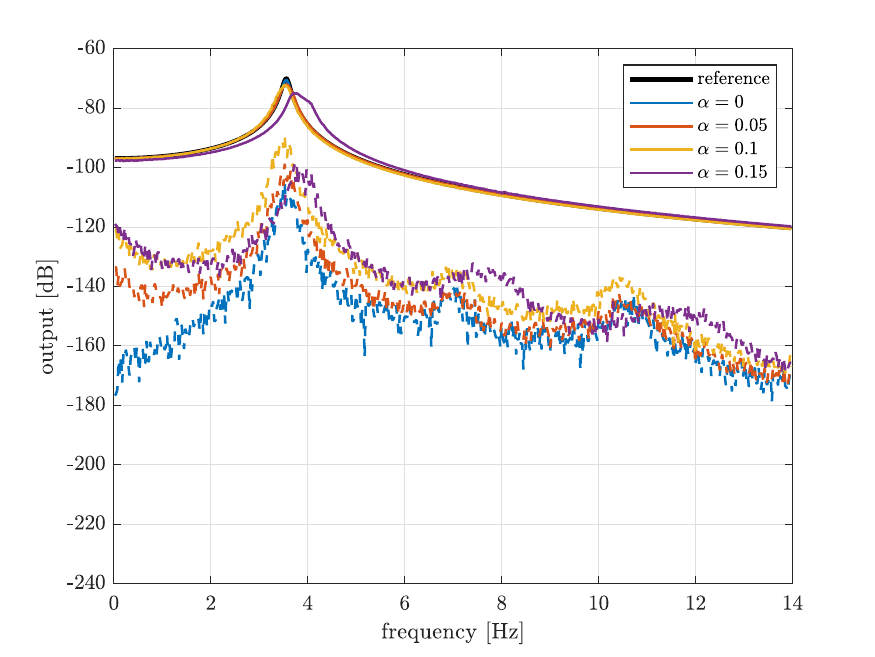}
      \caption{}
    \end{subfigure}
    \begin{subfigure}{.49\textwidth}
      \centering
      \includegraphics[width=\linewidth]{  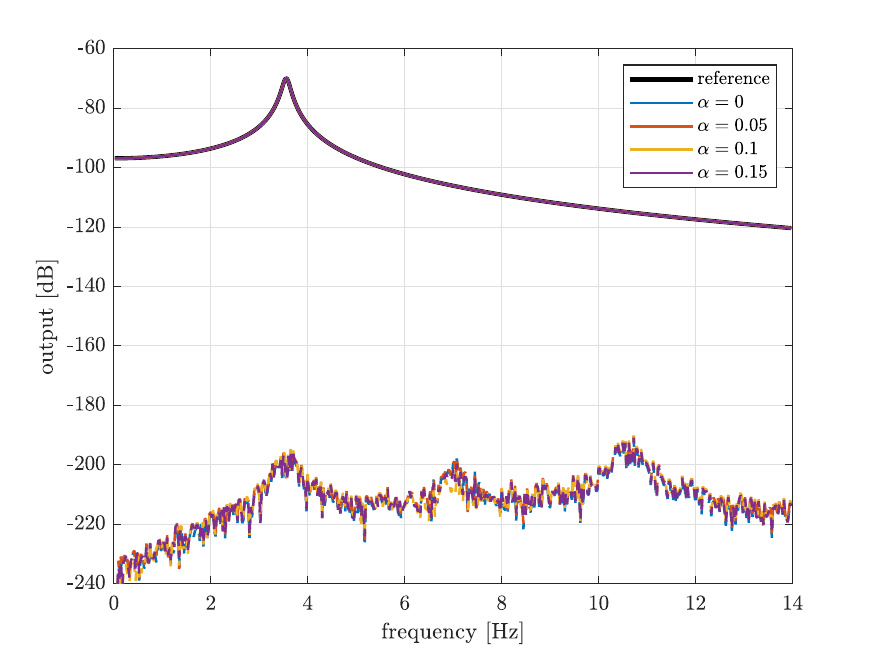}
      \caption{}
    \end{subfigure}
    \caption{\color{red}{Nonlinear distortion analysis after linearisation for various values of $\alpha$, for (a) classical feedback linearisation and (b) the proposed method.}}
    \label{fig:compare}
\end{figure}

\subsubsection{Robustness against unmodelled nonlinearities}
Fig. \ref{fig:nldafter4nq} displays the nonlinear distortion analysis when only the cubic nonlinearity is considered. This is achieved by setting the first column of the $\bm{E}$ matrix of the model in Table \ref{table:sys1} to zero. By doing so, the linear part of the model remains exactly the same as in the previous performance test, in this way allowing for a valid comparison between the two. As expected, the even nonlinearities are visibly more present when compared to Fig. \ref{fig:nldafter4}. In addition, almost no improvement regarding the suppression of the even nonlinearities is noticed when compared to the original system before linearisation in Fig. \ref{fig:nldbefore4}. Only around the resonance frequency, there is a reduction of approximately 15 dB. Nevertheless, the odd nonlinearities are all reduced to a satisfactory level, similarly to Fig. \ref{fig:nldafter4}.\\

\begin{figure}[!t]
    \centering
    \includegraphics[width=.65\textwidth]{  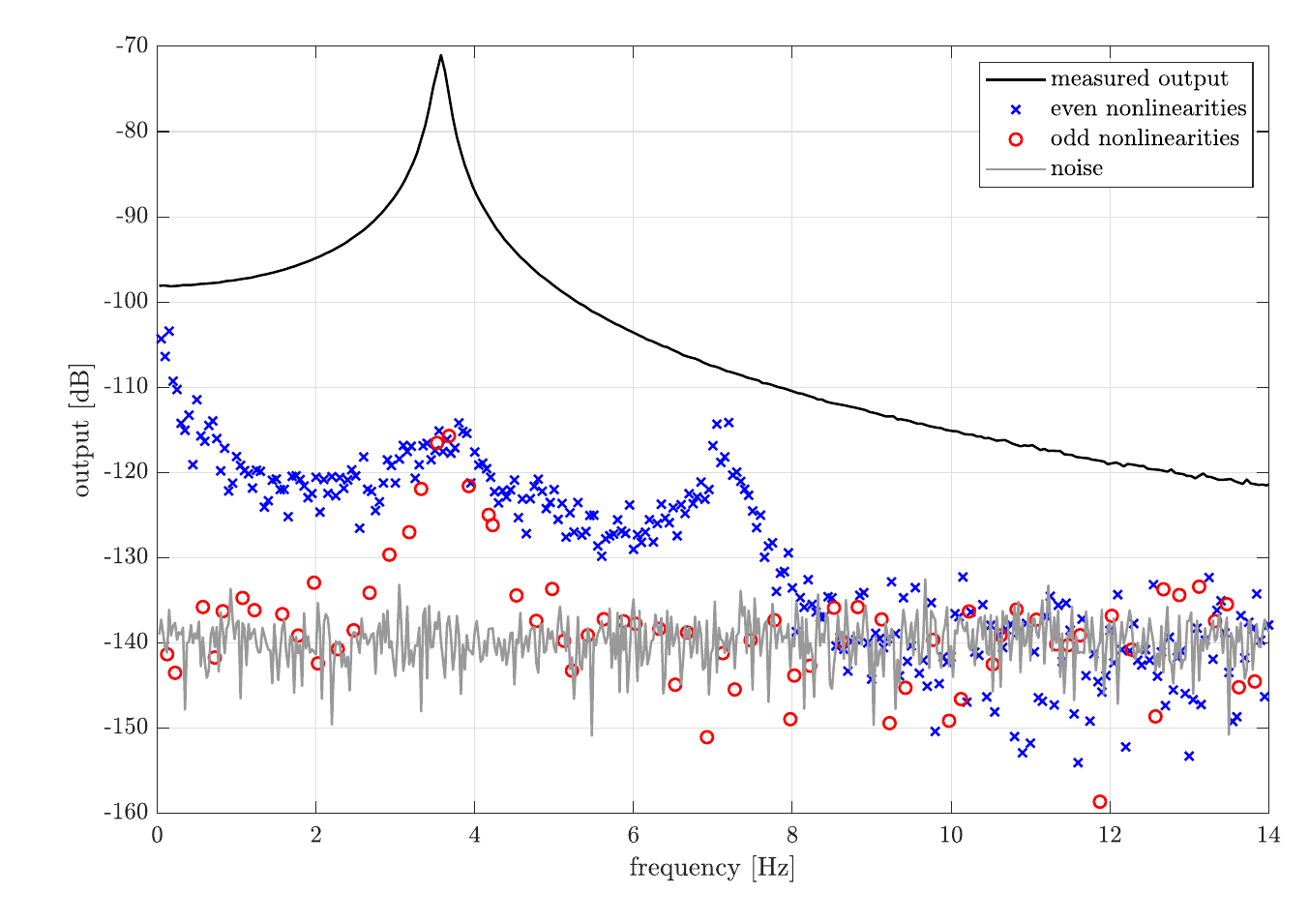}
    \caption{Nonlinear distortion analysis after linearisation in the presence of modelling errors.}
    \label{fig:nldafter4nq}
\end{figure}
\begin{figure}[!t]
    \centering
    \begin{subfigure}{.49\textwidth}
      \centering
      \includegraphics[width=\linewidth]{  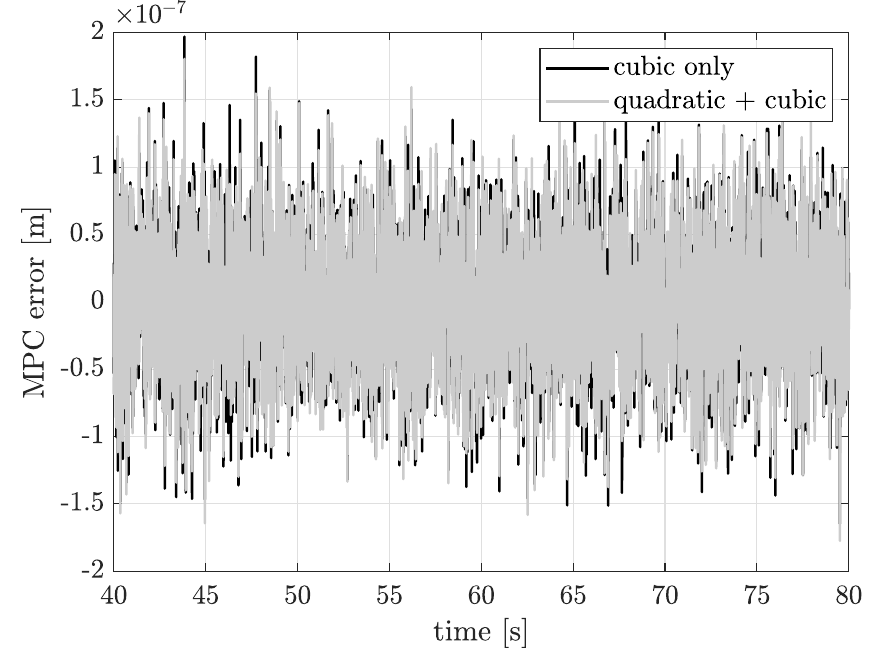}
      \caption{}
    \end{subfigure}
    \begin{subfigure}{.49\textwidth}
      \centering
      \includegraphics[width=\linewidth]{  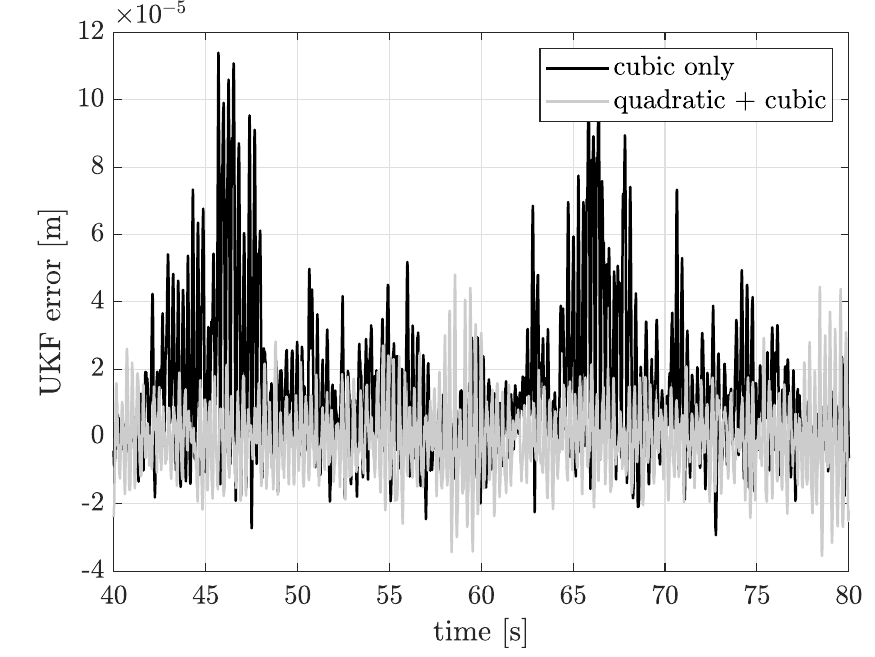}
      \caption{}
    \end{subfigure}
    \caption{(a) MPC and (b) UKF error plots for one period of the multisine. The errors for the complete (quadratic + cubic, in grey) and incomplete model (cubic only, in black) are shown. The MPC handles modelling errors very well due to its integral action property. On the contrary, the performance of the UKF significantly degrades when the quadratic nonlinearity is not modelled.}
    \label{fig:errorss}
\end{figure}

Fig. \ref{fig:errorss} allows for more insight in the observations drawn from the distortion analysis. Here, two different error signals for both the complete and incomplete models are shown; the left figure presents the tracking error, \textit{i.e.}, the filtered output minus the reference output, and the right figure shows the UKF error, \textit{i.e.}, the filtered output minus the true output, the true output being the noise-free output accessible in simulation. It can be seen that there is no noticeable difference between the tracking error signals. Specifically, this means that the controller is able to match the \textit{filtered} output and the reference output equally well for both the complete and incomplete models. Therefore, it can be concluded that the integral action property of the linearising controller performs properly. The UKF error plot explains the poor suppression of the even nonlinearities. An increase of the UKF error is noted for the incomplete model. Table \ref{table:robust} gives the RMS values of all the errors and the ratio between these errors and the measured output. In conclusion, the incomplete model is not able to provide an accurate estimate of the true output, explaining the poor reduction of the even nonlinearities in Fig. \ref{fig:nldafter4nq}.

\subsubsection{Robustness against extrapolation}

\begin{table}
\begin{center}
    \begin{tabular}{c||c|c|c}
         & measured output & tracking error & UKF error \\
         \hline \hline
        cubic only & $4.14\cdot10^{-4}$  & $2.27\cdot10^{-7}\;\;(0.0548\%)$ & $2.64\cdot10^{-5}\;\;(6.38\%)$ \\
        \hline
        quadratic + cubic & $3.82\cdot10^{-4}$ & $2.22\cdot10^{-7}\;\;(0.0581\%)$ & $1.14\cdot10^{-5}\;\;(2.98\%)$\\
        \hline
        extrapolated & $6.51\cdot10^{-4}$ & $2.36\cdot10^{-7}\;\;(0.0363\%)$ & $2.12\cdot10^{-5}\;\;(3.26\%)$\\
    \end{tabular}
\end{center}
\caption{RMS values (in meter) of the measured output, tracking error and UKF error for the complete and incomplete models (test 1), and the complete model when extrapolated (test 2). The percentages between brackets correspond to the ratios of the errors to the measured output.}
\label{table:robust}
\end{table}
\begin{figure}
    \centering
    \begin{subfigure}{.49\textwidth}
      \centering
      \includegraphics[width=\linewidth]{  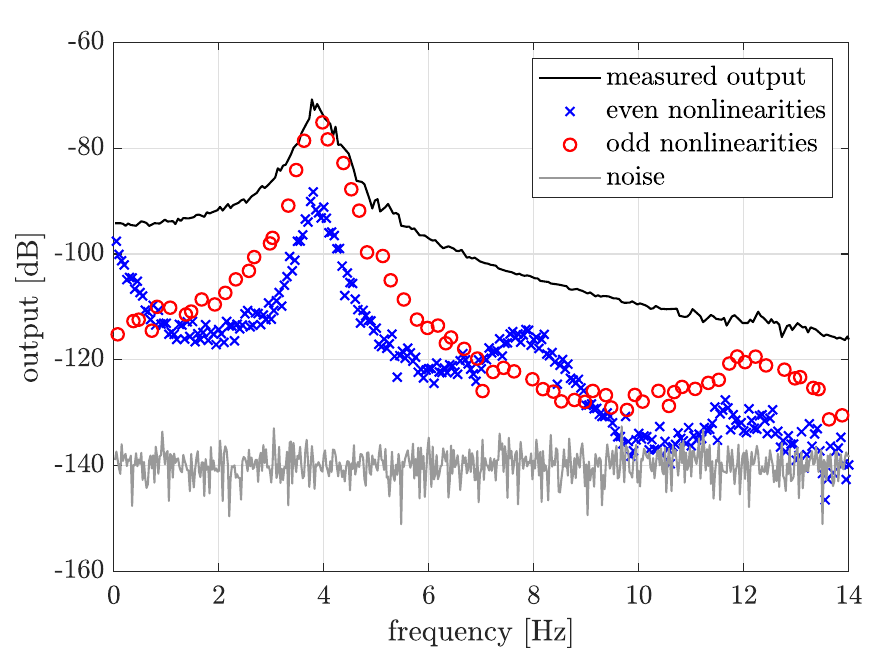}
      \caption{}
      \label{fig:NLDperfecta}
    \end{subfigure}
    \begin{subfigure}{.49\textwidth}
      \centering
      \includegraphics[width=\linewidth]{  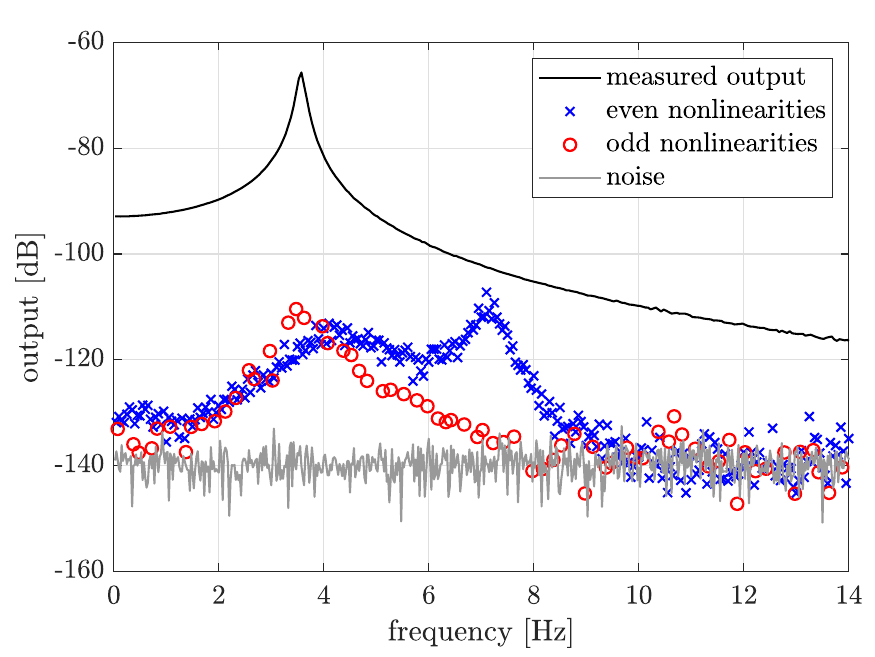}
      \caption{}
      \label{fig:NLDperfectb}
    \end{subfigure}
    \caption{Nonlinear distortion analysis (a) before and (b) after linearisation when excited with a random-phase multisine with an RMS amplitude of 0.22 N.}
    \label{fig:NLDperfect}
\end{figure}
% \begin{figure}
%     \centering
%     \begin{subfigure}{.49\textwidth}
%       \centering
%       \includegraphics[width=\linewidth]{  mpcerror2.pdf}
%       \caption{}
%     \end{subfigure}
%     \begin{subfigure}{.49\textwidth}
%       \centering
%       \includegraphics[width=\linewidth]{  ukferror2.pdf}
%       \caption{}
%     \end{subfigure}
%     \caption{(a) MPC and (b) UKF error plots for one period of a multisine with an RMS amplitude of 0.22 N.}
%     \label{fig:error2}
% \end{figure}

To challenge the extrapolation capability of the control architecture, a random-phase multisine with a RMS amplitude of 0.22 N is used, compared to 0.12 N in the previous sections. Fig. \ref{fig:NLDperfect} presents the nonlinear distortion analysis (left) before and (right) after linearisation. In the left subfigure, nonlinear distortions are observed to be markedly greater when compared to the analysis in Fig. \ref{fig:nldbefore4}. Nevertheless, a satisfactory linearisation is achieved in the right subplot, the results being similar to Fig. \ref{fig:nldafter4}, with the peaks around 3.8 and 7.6 Hz slightly amplified.\\

It is interesting to notice that the low-frequency performance is considerably better than in the case of the model without quadratic nonlinearity in Fig. \ref{fig:nldafter4nq}. The tracking and UKF errors are provided in Table \ref{table:robust}, from which can be seen that the magnitude of the tracking error is similar to the values reported in the previous robustness test, even though the RMS value of the measured output increased by more than 70 percent. This further confirms the excellent working of the integral action property. As expected, the magnitude of the UKF error also increased significantly. Yet, in percent, it compares to the full nonlinear model error. One may then conclude that the poor low-frequency performance in the previous robustness test is due to the nature of the UKF error in Fig. \ref{fig:errorss}. Indeed, there clearly exists an offset in the error signal. From a physical point of view, this offset is explainable since the quadratic nonlinearity is not modelled. In the frequency domain, an offset translates in a non-zero line at 0 Hz. It is therefore coherent that the low-frequency region is mostly affected by the UKF-induced offset.

    \section{Experimental validation using a prototype of a high-precision motion system} \label{sec:experiments}
\begin{figure}[!b]
    \centering
    \includegraphics[width=\textwidth]{  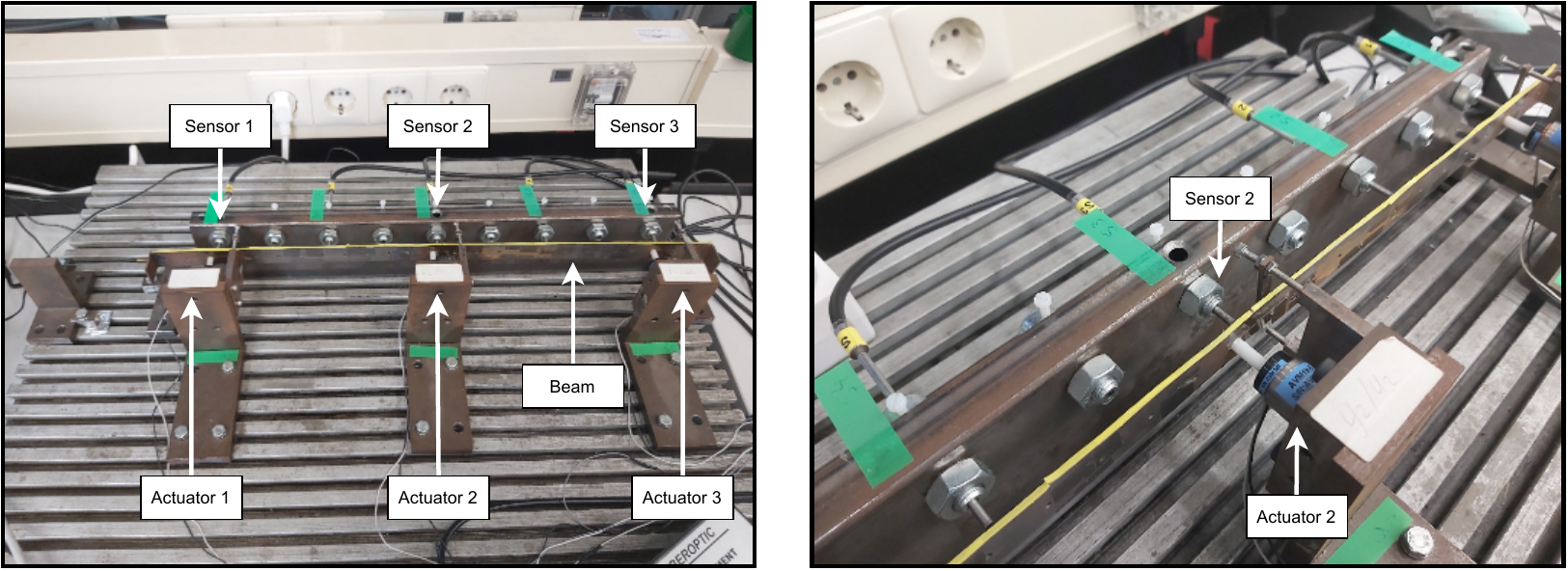}
    \caption{Pictures of the experimental flexible beam setup, equipped with 3 collocated sensor-actuator pairs. In this work, only the middle pair is exploited.}
    \label{fig:beam}
\end{figure}
    In this section, the proposed control framework is applied to an experimental prototype of a high-precision flexible positioning system, developed for evaluating control strategies. The system, shown in Fig. \ref{fig:beam}, consists of a lightweight flexible steel beam with dimensions $500\times20\times2$ mm. The boundary conditions of the beam are realised by means of leaf springs, which constrain four out of six degrees of freedom (DOFs). {Effectively, this means that the beam is only able to move in one translational and one rotational direction, both in the horizontal plane.} The setup is equipped with three collocated sensor-actuator pairs, operating at a sampling frequency of 4096 Hz. The actuators are current-driven voice-coil actuators, whereas the sensors are contactless fiberoptic displacement sensors with an approximate accuracy of 1 $\mu$m. Since the system has more actuators than DOFs, it is said to be over-actuated \cite{classens2021closed}. However, over-actuation is not considered in this work; in fact, only the middle sensor-actuator pair is exploited. Finally, a positive cubic stiffness, \textit{i.e.}, a hardening spring, is artificially implemented in the setup as is explained in Fig. \ref{fig:beamie}. The value of the spring constant is set to $k_c=2\cdot10^9$ V/m$^3$. By doing so, the obtained dynamics is dominated by the nonlinearity, in the sense that nonlinear distortions under a high-level multisine excitation are less than $10$ dB below the linear system dynamics in the first resonance region. 
\begin{figure}[!t]
    \centering
    \includegraphics{  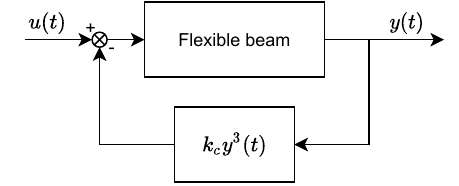}
    \caption{Block diagram of the experimental setup used for validating the proposed framework. {The input $u(t)$ is the voltage [V] over the voice-coil actuator; the output $y(t)$ is the measured displacement [m] of the beam.} A hardening spring is artificially created by feeding back the output sensor measurement, through a cubic function, to the input. The cubic spring constant is set to $k_c=2\cdot10^9$ V/m$^3$.}
    \label{fig:beamie}
\end{figure}
\begin{figure}[!t]
    \centering
    \includegraphics[width=.65\textwidth]{  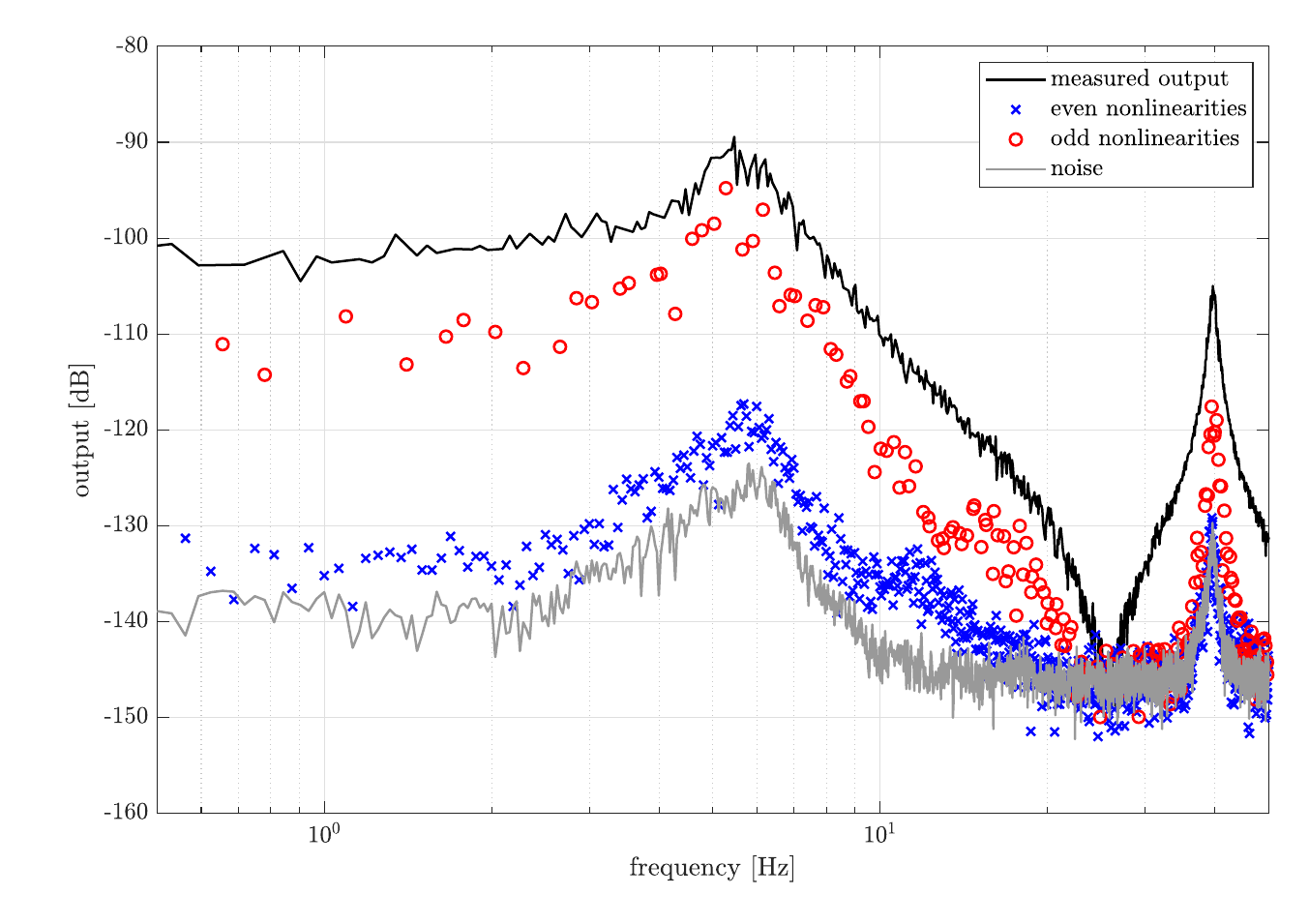}
    \caption{Nonlinear distortion analysis of the system with an artificial nonlinearity created following Fig. \ref{fig:beamie}.}
    \label{fig:nldbeforeexp}
\end{figure}

Fig. \ref{fig:nldbeforeexp} shows the nonlinear distortion analysis of the system with artificial nonlinearity. It is excited with 5 realisations of 4 periods of an odd random-phase multisine with a RMS amplitude of 0.015 V. For every group of 4 odd frequency lines, 1 line is randomly rejected. Moreover, 3280 points per period are considered at a sampling frequency of 102.4 Hz. In this figure, the resonance around 4 Hz corresponds to a rigid-body mode of the flexible beam, and the resonance around 40 Hz is its first bending mode. Over the entire depicted frequency range, the even distortions are close to the noise floor, while the odd distortions are never more than one order of magnitude smaller than the measured output.

\subsection{System identification} \label{sec:expsysid}
The identification data for the system is obtained by performing 10 realisations of 5 periods of a full random-phase multisine from 0 to 50 Hz and with a RMS amplitude of 0.015 V. The number of data points per period is 16384 at a sampling frequency of 1024 Hz. To ensure steady-state conditions, the first period of every realisation is rejected. One realisation is saved for validation.\\

Fig. \ref{fig:sysid2m} displays the fitting errors of the parameterised BLA (in orange) and of the full nonlinear model (in green), calculated on validation data. The nonlinear model simply embeds a $\zeta(y(t))=y^3(t)$ monomial; in consequence, the low magnitude even distortions in Fig. \ref{fig:nldbeforeexp} are not modelled as they are judged negligible. The left plot of Fig. \ref{fig:sysid2m} presents one time-domain period of the validation data with the corresponding linear and nonlinear modelling errors; similar results as in the numerical demonstration in Section \ref{sec:simulations} are obtained. The nonlinear model reduces the identification error significantly, from 57.4$\%$ to 7.34$\%$, when compared to the linear model. In the right subfigure, the validation data is presented in the frequency domain. Up to 15 Hz, the nonlinear error is consistently around 30 dB below the validation output. The parameter values of the obtained nonlinear state-space model are provided in Table \ref{table:sys3} of \ref{appendix:Models}.\\
\begin{figure}[!t]
    \centering
    \begin{subfigure}{.49\textwidth}
      \centering
      \includegraphics[width=\linewidth]{  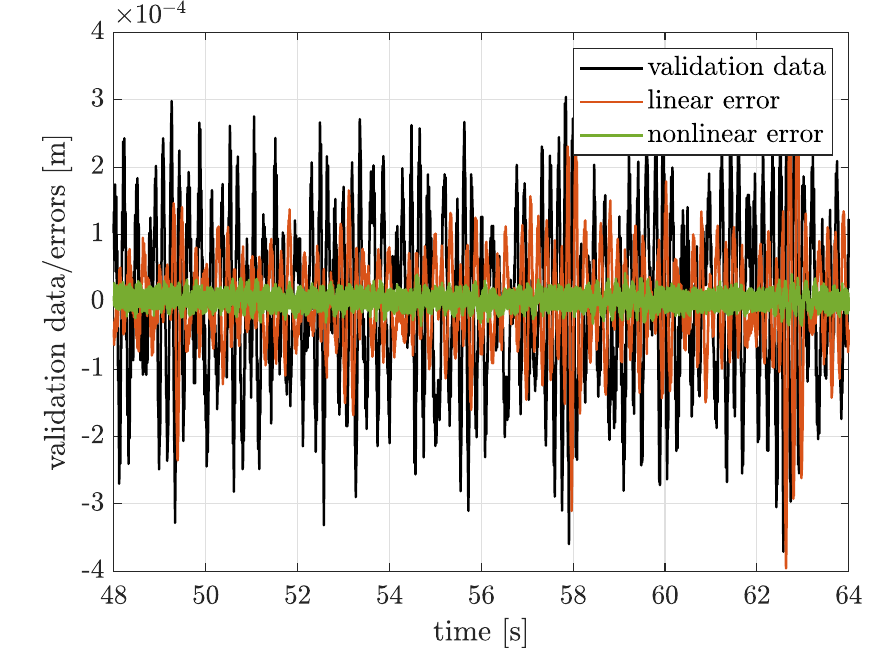}
      \caption{}
    \end{subfigure}
    \begin{subfigure}{.49\textwidth}
      \centering
      \includegraphics[width=\linewidth]{  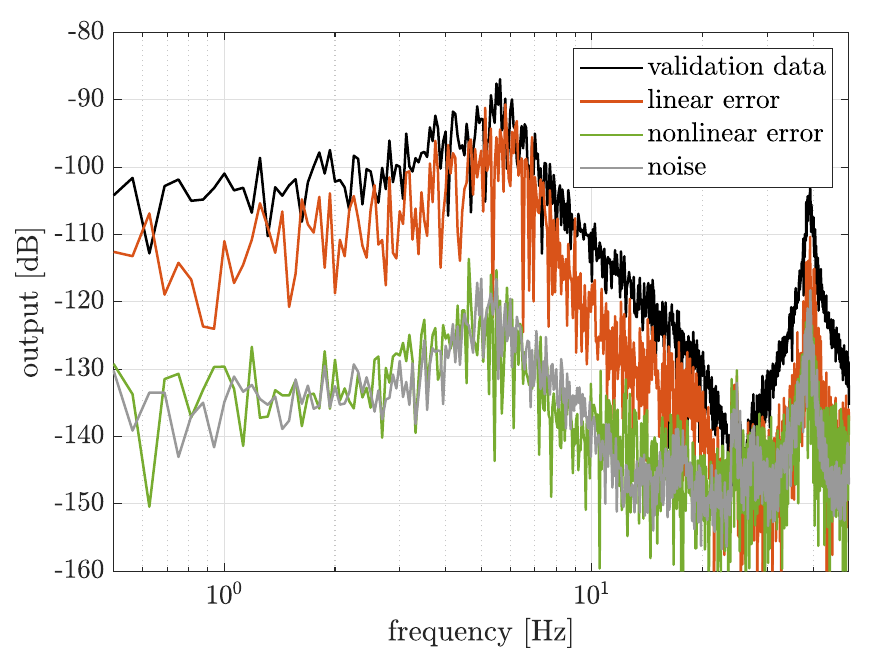}
      \caption{}
    \end{subfigure}
    \caption{(a) One period of the time-domain validation data and the associated identification errors of the parametric BLA (in orange) and of the nonlinear model (in green). The identification error is reduced from 57.4$\%$ to 7.34$\%$. (b) Corresponding frequency-domain validation plot.}
    \label{fig:sysid2m}
\end{figure}

\subsection{Linearisation performance}
In this section, the results of the linearising control framework for both a sine and a random multisine excitation are analysed. The control parameters are chosen following the guidelines discussed in Section \ref{sec:control}. Here, one must also consider the sampling time of the sensor-actuator pair. Namely, the nonlinear feedback structure of Fig. \ref{fig:beamie} represents the continuous plant. However, the sensor output is only subtracted from the input at the next time step, considering a sample rate of 4096 Hz. Therefore, the inner-loop sampling time should be sufficiently higher than 1/4096 s. While keeping this in mind, the outer-loop sample time is set to 1/102.4 s. In this way, a desired closed-loop bandwidth of around 20 Hz can be achieved by an outer-loop controller. Next, the inner-loop sampling time is set to 1/512 s, \textit{i.e.}, 5 times faster than the outer-loop while still sufficiently below 1/4096 s. The values of $Q$ and $R_{\Delta}$ are determined by performing noise-free experiments with the obtained model in simulation; they are set to $3\cdot10^8$ and 1, respectively. The values of the covariance matrices of the UKF are tuned on the setup itself. Again, the measurement noise variance is set to the mean value of the output noise variance of the nonlinear distortion analysis, \textit{i.e.}, $R_{UKF}=3.39\cdot10^{-15}$ m$^2$; the process noise covariance matrix is set to $Q_{UKF}=10R_{UKF}\bm{I}_4$.\\

Fig. \ref{fig:sineexpp} shows the plant input and output when excited by a sine signal. The sine has an amplitude of 0.02 V and a frequency of 2 Hz. The effect of the artificial nonlinear spring is clearly visible in the output plot. The IO behaviour of the linearised plant to the same outer-loop sine excitation is shown in Fig. \ref{fig:sinccoon}. The linearising controller satisfactorily tracks its reference down to an error of $1.05\%$. It can be observed that, in order to track the linear output, the peak amplitudes of the plant input are more than a factor 10 higher than that of the outer-loop sine input. This high amplitude is required to compensate for the severe hardening effect of the spring. Note that, in practical cases, it is more rational to adapt the amplitude of the outer-loop input such that an output is reached that matches the amplitude of the original nonlinear output (the one of Fig. \ref{fig:sineexpp2}); in that case, the required control input amplitudes would be significantly less severe.\\
\begin{figure}[!t]
    \centering
    \begin{subfigure}{.49\textwidth}
      \centering
      \includegraphics[width=\linewidth]{  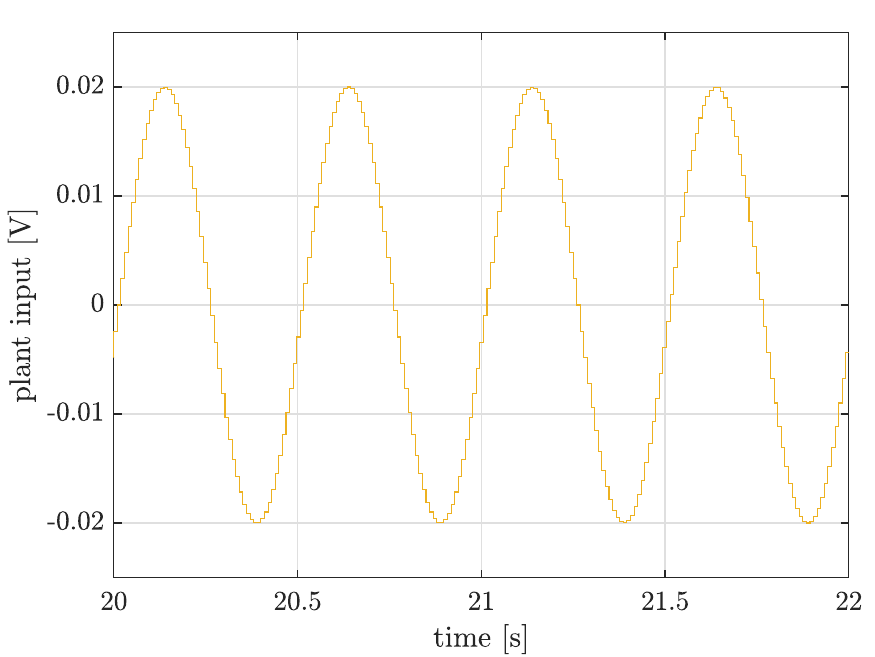}
      \caption{}
    \end{subfigure}
    \begin{subfigure}{.49\textwidth}
      \centering
      \includegraphics[width=\linewidth]{  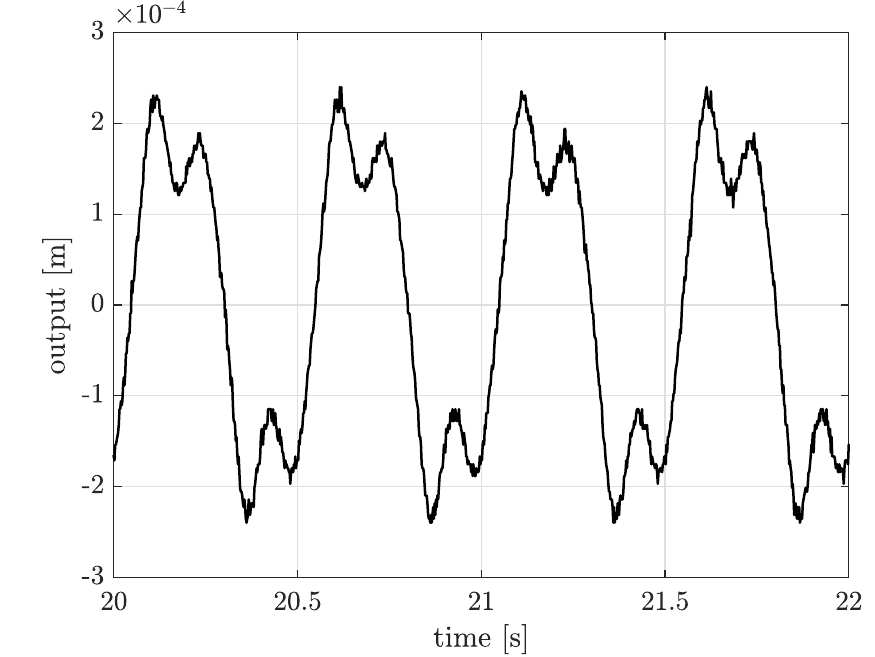}
      \caption{}
      \label{fig:sineexpp2}
    \end{subfigure}
    \caption{IO data when the original plant is excited by a sine signal with an amplitude of 0.02 V and a frequency of 2 Hz }
    \label{fig:sineexpp}
\end{figure}
\begin{figure}[!t]
    \centering
    \begin{subfigure}{.49\textwidth}
      \centering
      \includegraphics[width=\linewidth]{  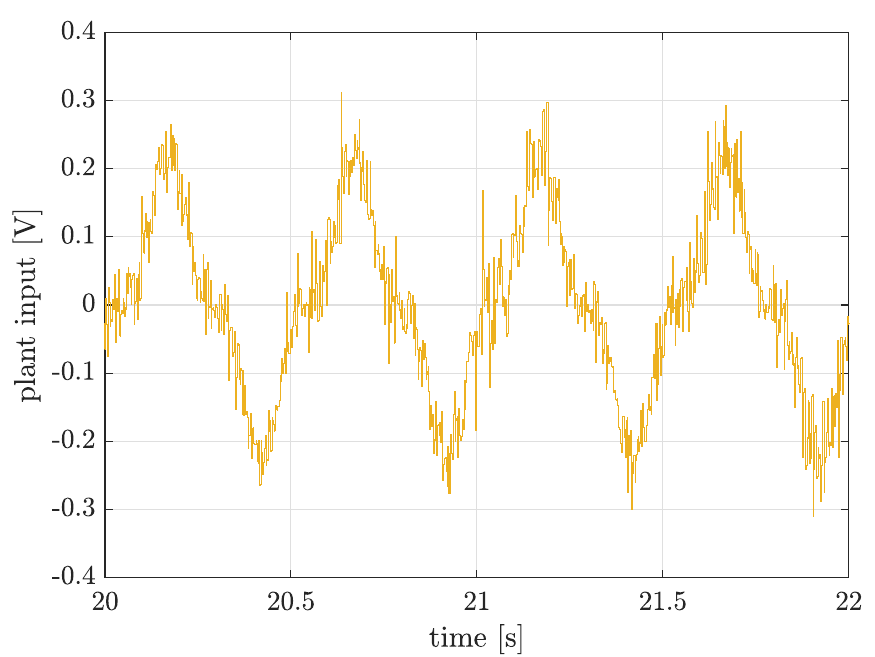}
      \caption{}
    \end{subfigure}
    \begin{subfigure}{.49\textwidth}
      \centering
      \includegraphics[width=\linewidth]{  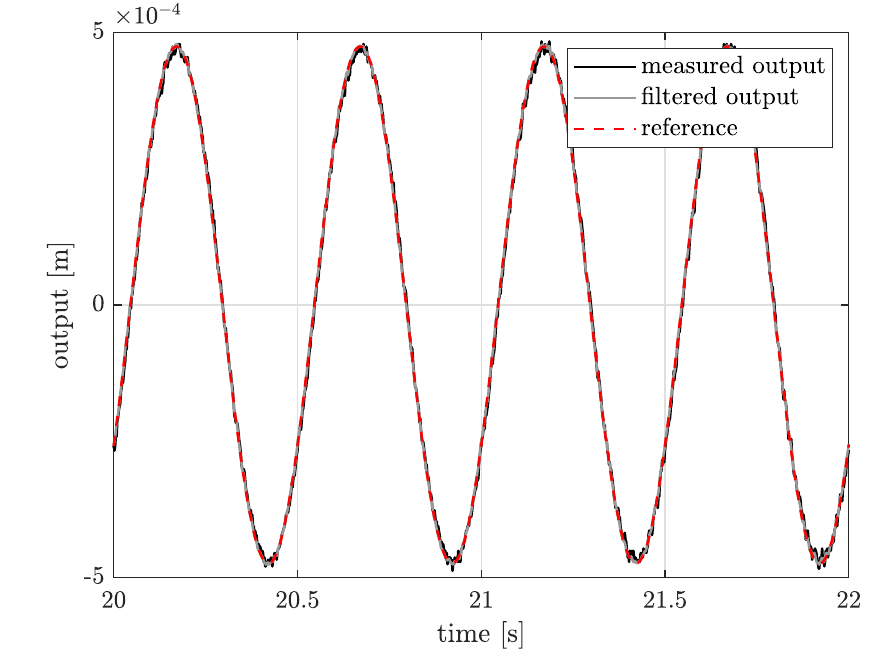}
      \caption{}
      \label{fig:outputtrack}
    \end{subfigure}
    \caption{IO data when the controlled plant is excited with the outer-loop sine signal of Fig. \ref{fig:sineexpp}. Compared to Fig. \ref{fig:sineexpp}, a significant increase of the plant input amplitude is required to counterbalance the severe hardening effect of the artificial spring.}
\label{fig:sinccoon}
\end{figure}

\begin{figure}[!t]
    \centering
    \includegraphics[width=.65\textwidth]{  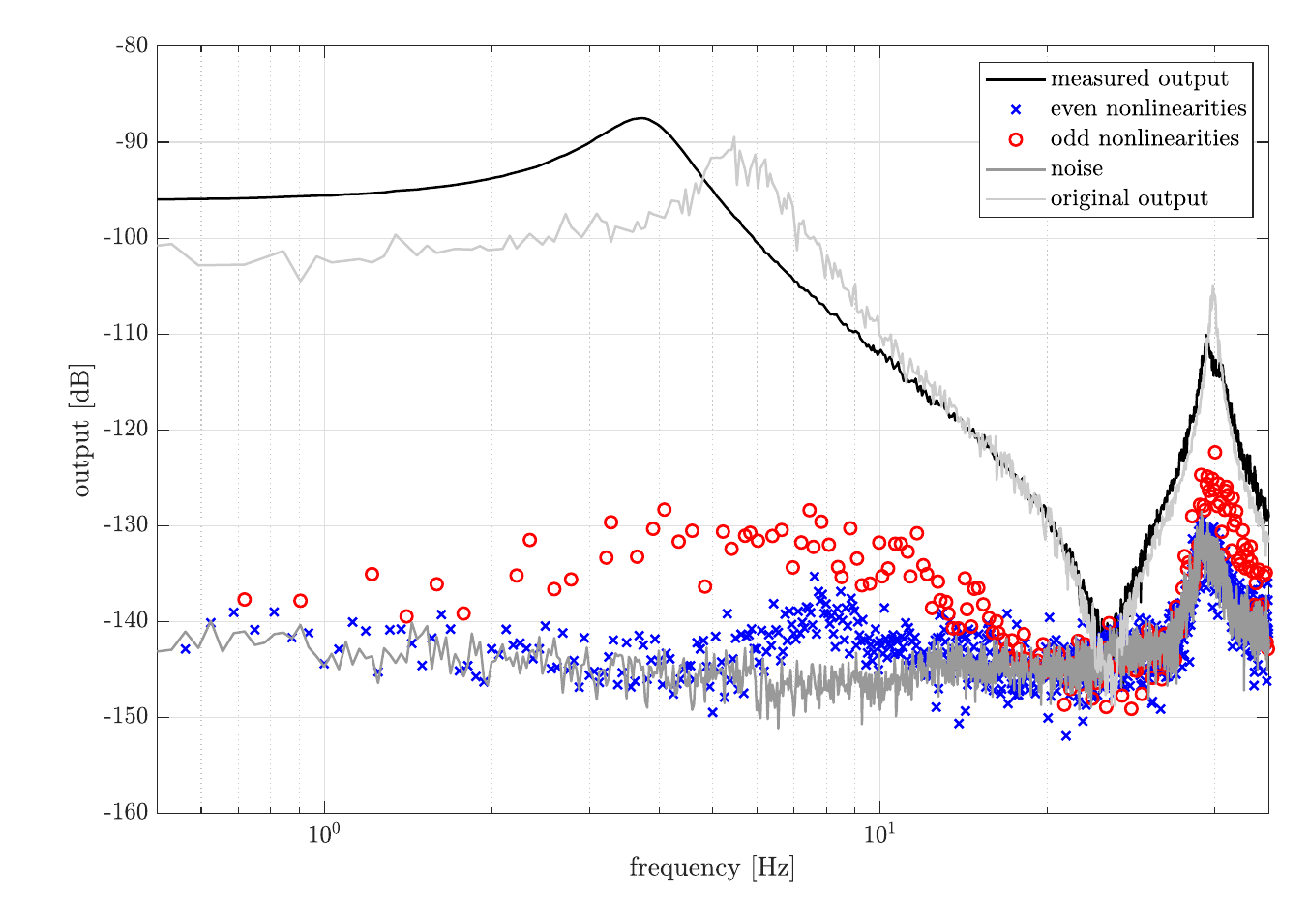}
    \caption{Nonlinear distortion analysis after linearisation, where the odd nonlinearities are significantly suppressed over the primary resonance.}
    \label{fig:nldafterb}
\end{figure}
Fig. \ref{fig:nldafterb} shows the nonlinear distortion analysis after linearisation. Excellent linearisation is achieved in the first resonance peak range; odd and even distortions are therein located within at most 10 dB of the noise floor. In more details, the odd distortions are eliminated by more than 30 dB, and the even distortions are completely suppressed, even though a cubic nonlinearity only is modelled. As expected, there is no significant performance improvement around the first flexible mode of the beam. As a final performance indicator, Fig. \ref{fig:nldafterb} plots the output of the plant before linearisation (in light grey), \textit{i.e.}, the output data in Fig. \ref{fig:nldbeforeexp}. It is, in this figure, clearly visible that the effect of the hardening spring is eliminated; the resonance peak is shifted back to lower frequencies and the static output value at 0 Hz increases. The colouring of the noise observed in Fig. \ref{fig:nldbeforeexp} is also whitened.

    \section{Conclusion} \label{sec:conclusion}
In this paper, a data-driven approach towards feedback linearisation of nonlinear mechanical systems was presented. The introduced framework is especially designed for nonlinear mechanical systems, but is theoretically applicable to any plant that can be modelled as a linear, time-invariant system with static output nonlinearities in the feedback path. An optimisation-free model-based reference tracking methodology was proposed that eliminates the undesirable nonlinearities from the input-output response, while at the same time preserves the important linear dynamics. 

The presented work incorporated controller design and implementation, as well as an intuitive nonparametric nonlinearity analysis. Robustness tests in a realistic simulation environment showed that the linearising controller achieves very low tracking errors when severe modelling errors are present, or when the model is extrapolated outside its fitting region. Here, linearising performance is mainly hampered by the unscented Kalman filter, which relies significantly more on an accurate model. Experimental results on a prototype of a highly nonlinear positioning system confirmed the excellent working of the proposed framework. Suggestions for future research are the extension of the proposed method to the multi-input multi-output case, and to systems with input and state nonlinearities.

    \section*{Acknowledgements}
    This research did not receive any specific grant from funding agencies in the public, commercial, or not-for-profit sectors.
    
    \bibliographystyle{IEEEtran}
    \bibliography{Refs}
    \appendix
    \section{Derivation of the MPC prediction matrices} \label{appendix:MPC}
This appendix concerns the derivation of the MPC prediction matrices, which are used to describe the evolution of the states and output over the prediction horizon. Remember the relations $t(0|k)=t(k) \in \mathbb{R}_{\geq0}$, $\olsi{x}(0|k)=\olsi{x}(k) \in \mathbb{R}^n$, $\Delta u(0|k)=\Delta {u}(k) \in \mathbb{R}$ and $\Delta \zeta(y(0|k))=\Delta \zeta(y(k)) \in \mathbb{R}^s$. Based on the augmented state-space (\ref{eq:augmstatespace}), the $N_p$ future predictions of the output at time $t(k)$, \textit{i.e.},
\begin{equation}
    Y_k = \left[y(i+1|k), \ldots, y(i+N_p| k)\right]^{\top},
\end{equation}
are given by the following relations
\begin{align}
    \begin{split}
    &\begin{cases}
    \olsi{x}(1|k) \hphantom{,.} &= \olsi{\bm{A}}\olsi{x}(k) + \olsi{\bm{B}}\Delta u(k) + \olsi{\bm{E}}\Delta \zeta(y(k))\\
    y(1|k) &= \olsi{\bm{C}}\olsi{x}(1|k)\\
           &= \olsi{\bm{C}}\olsi{\bm{A}}\olsi{x}(k)+\olsi{\bm{C}}\olsi{\bm{B}}\Delta u(k) +\olsi{\bm{C}}\olsi{\bm{E}}\Delta \zeta(y(k)),
    \end{cases}\\
    &\begin{cases}
    \olsi{x}(2|k) \hphantom{,.} &= \olsi{\bm{A}}\olsi{x}(1|k) + \olsi{\bm{B}}\Delta u(1|k) + \olsi{\bm{E}}\Delta \zeta(y(1|k))\\
                  &= \olsi{\bm{A}}^2\olsi{x}(k) + \olsi{\bm{A}}\olsi{\bm{B}}\Delta u(k) + \olsi{\bm{A}}\olsi{\bm{E}}\Delta \zeta(y(k)) + \olsi{\bm{B}}\Delta u(1|k) + \olsi{\bm{E}}\Delta \zeta(y(1|k))\\
    y(2|k) &= \olsi{\bm{C}}\olsi{x}(2|k)\\
           &= \olsi{\bm{C}}\olsi{\bm{A}}^2\olsi{x}(k) + \olsi{\bm{C}}\olsi{\bm{A}}\olsi{\bm{B}}\Delta u(k) + \olsi{\bm{C}}\olsi{\bm{A}}\olsi{\bm{E}}\Delta \zeta(y(k)) + \olsi{\bm{C}}\olsi{\bm{B}}\Delta u(1|k)\\
           &\quad + \olsi{\bm{C}}\olsi{\bm{E}}\Delta \zeta(y(1|k)),
    \end{cases}\\
    &\mathrel{\makebox[\widthof{==============,}]{\vdots}}\\
    &\begin{cases}
    \olsi{x}(N_p|k) &= \olsi{\bm{A}}\olsi{x}(N_p-1|k) + \olsi{\bm{B}}\Delta u(N_p-1|k) + \olsi{\bm{E}}\Delta \zeta(y(N_p-1|k))\\
                    &= \olsi{\bm{A}}^{N_p}\olsi{x}(k) + \olsi{\bm{A}}^{N_p-1}\olsi{\bm{B}}\Delta u(k) + \olsi{\bm{A}}^{N_p-1}\olsi{\bm{E}}\Delta \zeta(y(k))\\
                    &\quad + \ldots + \olsi{\bm{A}}\olsi{\bm{B}}\Delta u(N_p-2|k) + \olsi{\bm{A}}\olsi{\bm{E}}\Delta \zeta(y(N_p-2|k)) + \olsi{\bm{B}}\Delta u(N_p-1|k)\\
                    &\quad+ \olsi{\bm{E}}\Delta \zeta(y(N_p-1|k))\\
    y(N_p|k) &= \olsi{\bm{C}}\olsi{x}(N_p|k)\\
             &= \olsi{\bm{C}}\olsi{\bm{A}}^{N_p}\olsi{x}(k) + \olsi{\bm{C}}\olsi{\bm{A}}^{N_p-1}\olsi{\bm{B}}\Delta u(k) + \olsi{\bm{C}}\olsi{\bm{A}}^{N_p-1}\olsi{\bm{E}}\Delta \zeta(y(k))\\
             &\quad + \ldots + \olsi{\bm{C}}\olsi{\bm{A}}\olsi{\bm{B}}\Delta u(N_p-2|k) + \olsi{\bm{C}}\olsi{\bm{A}}\olsi{\bm{E}}\Delta \zeta(y(N_p-2|k))\\
             &\quad + \olsi{\bm{C}}\olsi{\bm{B}}\Delta u(N_p-1|k) +
             \olsi{\bm{C}}\olsi{\bm{E}}\Delta \zeta(y(N_p-1|k)),
    \end{cases}
    \end{split}
    \label{eq:Yklong}
\end{align}
which can be cast into a more compact form as follows
\begin{equation}
    Y_k = S_x\olsi{x}(k) + S_u\Delta U_k + S_g \Delta G_k.
\end{equation}
where
\begin{align}
    S_x = \left[\begin{array}{c}
    \olsi{\bm{C}}\olsi{\bm{A}}\\
    \olsi{\bm{C}}\olsi{\bm{A}}^2\\
    \vdots\\
    \olsi{\bm{C}}\olsi{\bm{A}}^{N_p}
    \end{array}\right];
    \;\,
    S_u = \left[\begin{array}{cccc}
    \olsi{\bm{C}}\olsi{\bm{B}} & 0 & \cdots & 0 \\
    \olsi{\bm{C}}\olsi{\bm{A}}\olsi{\bm{B}} & \olsi{\bm{C}}\olsi{\bm{B}} & \cdots & 0 \\
    \vdots & \vdots & \ddots & \vdots \\
    \olsi{\bm{C}}\olsi{\bm{A}}^{N_p-1}\olsi{\bm{B}} & \olsi{\bm{C}}\olsi{\bm{A}}^{N_p-2}\olsi{\bm{B}} & \cdots & \olsi{\bm{C}}\olsi{\bm{B}}
    \end{array}\right];
    \;\,
    S_g = \left[\begin{array}{cccc}
    \olsi{\bm{C}}\olsi{\bm{E}} & \bm{0}_{1\times s}  & \cdots & \bm{0}_{1\times s} \\
    \olsi{\bm{C}}\olsi{\bm{A}}\olsi{\bm{E}} & \olsi{\bm{C}}\olsi{\bm{E}} & \cdots & \bm{0}_{1\times s}\\
    \vdots & \vdots & \ddots & \vdots \\
    \olsi{\bm{C}}\olsi{\bm{A}}^{N_p-1}\olsi{\bm{E}} & \olsi{\bm{C}}\olsi{\bm{A}}^{N_p-2}\olsi{\bm{E}} & \cdots & \olsi{\bm{C}}\olsi{\bm{E}}
    \end{array}\right],
\end{align}
and
\begin{align}
    \Delta U_k = \left[\begin{array}{c}
    \Delta u(k)\\
    \Delta u(1|k)\\
    \vdots\\
    \Delta u(N_p-1|k)\\
    \end{array}\right];
    \qquad
    \Delta G_k = \left[\begin{array}{c}
    \Delta \zeta(y(k))\\
    \Delta \zeta(y(1|k))\\
    \vdots\\
    \Delta \zeta(y(N_p-1|k))\\
    \end{array}\right].
\end{align}
Ultimately, $\Delta U_k$ is the vector to be optimised while $\olsi{x}(k)$ is constructed from output measurements and $\Delta G_k$ is known (or estimated, technically) in advance. 
The obtained relations can be generalised to any time $t(i|k)$ and for any length of the prediction horizon $N_p$.
    \section{Nonlinear state-space parameter values} \label{appendix:Models}
This appendix provides the parameter values of the two nonlinear state-space models derived in the paper. {The models are obtained by following the four-step procedure in Section \ref{sec:nlsysid}}. Table \ref{table:sys1} corresponds to the simulation experiments of Section \ref{sec:simulations}, and Table \ref{table:sys3} details the model of the experimental setup in Section \ref{sec:experiments}.
\begin{table}[H]
\begin{center}
    \begin{tabular}{c|c||c|c||c|c}
         Parameter & Value & Parameter & Value & Parameter & Value \\
         \hline 
         $\bm{A}(1,1)$ & $9.992\cdot10^{-1}$ & $\bm{B}(1,1)$ & $-2.468\cdot10^{-3}$ & $\bm{E}(1,1)$ & $1.326\cdot10^{2}$\\
         $\bm{A}(2,1)$ & $-2.070\cdot10^{-2}$ & $\bm{B}(2,1)$ & $2.916\cdot10^{-4}$ & $\bm{E}(2,1)$ & $7.221$\\
         $\bm{A}(1,2)$ & $2.428\cdot10^{-2}$& $\bm{C}(1,1)$ & $2.467\cdot10^{-3}$ & $\bm{E}(1,2)$ & $2.598\cdot10^{5}$\\
         $\bm{A}(2,2)$ & $9.994\cdot10^{-1}$& $\bm{C}(1,2)$ & $1.854\cdot10^{-2}$ & $\bm{E}(2,2)$ & $-4.306\cdot10^{4}$\\
        \hline
    \end{tabular}
\end{center}
\caption{Second-order discrete-time state-space parameters of the full simulation model ($t_s=1/1000$ s).}
\label{table:sys1}
\end{table}

\begin{table}[H]
\begin{center}
    \begin{tabular}{c|c||c|c||c|c}
         Parameter & Value & Parameter & Value & Parameter & Value \\
         \hline 
         $\bm{A}(1,1)$ & $9.739\cdot10^{-1}$ & $\bm{A}(3,3)$ & $1.001$ & $\bm{C}(1,1)$ & $-1.297\cdot10^{-2}$\\
         $\bm{A}(2,1)$ & $-2.270\cdot10^{-1}$ & $\bm{A}(4,3)$ & $1.393\cdot10^{-2}$ & $\bm{C}(1,2)$ & $1.534\cdot10^{-3}$\\
         $\bm{A}(3,1)$ & $-4.9715\cdot10^{-4}$& $\bm{A}(1,4)$ & $9.026\cdot10^{-4}$ & $\bm{C}(1,3)$ & $-5.293\cdot10^{-2}$\\
         $\bm{A}(4,1)$ & $-7.984\cdot10^{-4}$& $\bm{A}(2,4)$ & $-2.339\cdot10^{-2}$ & $\bm{C}(1,4)$ & $9.680\cdot10^{-3}$\\ 
         $\bm{A}(1,2)$ & $2.421\cdot10^{-1}$& $\bm{A}(3,4)$ & $-3.508\cdot10^{-2}$ & $\bm{E}(1,1)$ & $2.245\cdot10^{6}$\\
         $\bm{A}(2,2)$ & $9.631\cdot10^{-1}$ & $\bm{A}(4,4)$ & $9.913\cdot10^{-1}$ & $\bm{E}(2,1)$ & $7.039\cdot10^{6}$\\
         $\bm{A}(3,2)$ & $5.329\cdot10^{-3}$ & $\bm{B}(1,1)$ & $1.287\cdot10^{-3}$ & $\bm{E}(3,1)$ & $-2.326\cdot10^{6}$\\
         $\bm{A}(4,2)$ & $-4.458\cdot10^{-3}$& $\bm{B}(2,1)$ & $-3.121\cdot10^{-3}$ & $\bm{E}(4,1)$ & $-1.112\cdot10^{7}$\\
         $\bm{A}(1,3)$ & $-7.025\cdot10^{-3}$& $\bm{B}(3,1)$ & $1.750\cdot10^{-3}$ & \\ 
         $\bm{A}(2,3)$ & $-8.304\cdot10^{-3}$& $\bm{B}(4,1)$ & $5.684\cdot10^{-3}$ &  &\\
        \hline
    \end{tabular}
\end{center}
\caption{Fourth-order discrete-time state-space parameters of the experimental setup ($t_s=1/1024$ s).}
\label{table:sys3}
\end{table}

\end{document}